\documentclass[aps,superscriptaddress,nofootinbib,notitlepage]{revtex4-1}
\usepackage{amsmath}
\usepackage{epsfig}
\usepackage{hyperref}
\usepackage{url}
\def\bea#1\eea{\begin{align}#1\end{align}} 
\newcommand{\nnu}{\nonumber\\}
\newcommand{\bef}{\begin{figure}[htb]\centering}
\newcommand{\eef}{\end{figure}}

\begin{document}
\title{QCD evolution of the Sivers asymmetry}

\date{\today}

\author{Miguel G. Echevarria}
\email{m.g.echevarria@nikhef.nl}
\affiliation{Nikhef Theory Group, 
                  Science Park 105, 
                  1098 XG Amsterdam, The Netherlands}

\author{Ahmad Idilbi}
\email{aui13@psu.edu}
\affiliation{Department of Physics, 
                   Pennsylvania State University,
                   University Park, PA 16802, USA}

\author{Zhong-Bo Kang}
\email{zkang@lanl.gov}
\affiliation{Theoretical Division,
                   Los Alamos National Laboratory,
                   Los Alamos, NM 87545, USA}
                  
\author{Ivan Vitev}
\email{ivitev@lanl.gov}                   
\affiliation{Theoretical Division,
                   Los Alamos National Laboratory,
                   Los Alamos, NM 87545, USA}

\begin{abstract}
We study the QCD evolution of the Sivers effect in both semi-inclusive deep inelastic scattering (SIDIS) 
and Drell-Yan production (DY). 
We pay close attention to the non-perturbative spin-independent Sudakov 
factor in the evolution formalism and find a universal form which can describe reasonably well the experimental data on the transverse momentum distributions in SIDIS, DY lepton pair and $W/Z$ production. 
With this Sudakov factor at hand, we perform a global fitting of all the experimental data on the Sivers asymmetry in SIDIS from HERMES, COMPASS and Jefferson Lab. 
We then make predictions for the Sivers asymmetry in DY lepton pair and $W$ production that can be compared to the future experimental measurements to  
test the sign change of the Sivers functions between SIDIS and DY processes and constrain the sea quark Sivers functions. 

\end{abstract}

\maketitle

\section{Introduction}

In recent years, transverse spin physics has become one of the most active areas of high energy 
hadron physics research.  In particular, the experimental study and theoretical understanding 
of single transverse spin asymmetries has  resulted in a much deeper understanding of the nucleon 
structure~\cite{Boer:2011fh,Accardi:2012qut,Anselmino:2011ay,Aschenauer:2013woa,Peng:2014hta}.  
It has been realized that these observables can provide information on the parton's intrinsic transverse 
motion, which presents a path to three-dimensional proton tomography. The information about the three-dimensional 
proton structure is encoded in the so-called transverse 
momentum dependent  distribution functions (TMDs), which provide a new domain to study the strong 
interaction dynamics. They also open a new window to study the validity of  QCD factorization theorems and the
universality of the associated TMD parton distribution functions (TMDPDFs) and/or fragmentation functions 
(TMDFFs)~\cite{Brodsky:2002cx,Collins:2002kn,Boer:2003cm,Metz:2002iz,Kang:2011hk,Gamberg:2013kla,Gamberg:2012iq,Kang:2012xf,Metz:2012ui}. 

One of the most studied asymmetries has been the Sivers effect. It originates from a 
special TMD called the Sivers function~\cite{Sivers:1989cc}, which represents a distribution of 
unpolarized partons inside a transversely polarized proton through a correlation between the 
parton's transverse momentum and the proton polarization vector. The Sivers effect has gathered 
a lot of attention largely because of its unique property: the Sivers function is not exactly 
universal, instead, it exhibits  time-reversal modified 
universality~\cite{Brodsky:2002cx,Collins:2002kn,Boer:2003cm,Kang:2011hk,Kang:2009bp}. Based on parity and time-reversal invariance of QCD, it was shown that the quark Sivers function in semi-inclusive 
deep inelastic scattering (SIDIS) and those in the Drell-Yan (DY) process are equal in magnitude 
and  opposite in sign to each other. This sign change of the Sivers functions between SIDIS and DY 
is one of the most important predictions in the transverse spin physics and provides a critical test 
of the QCD factorization formalism and our understanding of  spin asymmetries. 

The Sivers asymmetry has been measured in the SIDIS process by HERMES~\cite{Airapetian:2009ae}, 
COMPASS~\cite{Alekseev:2008aa,Adolph:2012sp}, and Jefferson Lab (JLab)~\cite{Qian:2011py} experiments. 
Future measurements of the Sivers asymmetry in DY production have been 
planned~\cite{Aschenauer:2013woa,DY-compass-exp, DY-fermi-beam, DY-fermi-target, DY-rhic-exp} to 
verify the expected sign change. In anticipation 
of these new results, we need reliable predictions for the Sivers asymmetry in different processes. 
It is important to keep in mind that the Sivers asymmetry  was measured in SIDIS for typical 
momentum scales $Q \sim 1 - 3$ GeV, while for the DY-type processes it will be measured at 
much larger momentum scales $Q \sim 4 - 90$ GeV. Any reliable predictions will certainly have to 
include a correct understanding of the $Q$-dependence of the Sivers asymmetry. 
In other words, we 
have to properly include its energy evolution~\cite{collins-book,Kang:2011mr,Aybat:2011zv,Aybat:2011ge,Echevarria:2012pw,Sun:2013dya}. 

QCD evolution equations for the TMDs have been derived using different 
approaches~\cite{collins-book,Aybat:2011zv,Aybat:2011ge,Echevarria:2012pw},
and they are consistent with each other {\it perturbatively}. For QCD evolution equations 
of the associated spin-dependent collinear PDFs and/or FFs, see 
Refs.~\cite{Kang:2008ey,Zhou:2008mz,Vogelsang:2009pj,Braun:2009mi,Kang:2010xv,Schafer:2012ra,Kang:2012em,Ma:2012xn}. 
One of the difficulties related to the QCD evolution of  TMDs lies in the fact that
the complete evolution formalism  contains both perturbative and non-pertubative 
parts~\cite{Collins:1984kg,Qiu:2000ga,Landry:2002ix,Guzzi:2013aja,Nadolsky:2000ky,Aidala:2014hva}. Because of this, 
the evolved asymmetries in  phenomenological applications can be quite different depending on the 
treatment of the non-perturbative part~\cite{Aybat:2011ta,Anselmino:2012aa,Boer:2013zca,Sun:2013hua} 
even though the perturbative evolution kernel is exactly the same. The non-perturbative part should 
be universal and  extracted from the experimental data and in this paper we  pay close attention 
to its role in the evolution kernel. Thus, we first concentrate on the spin-averaged 
differential cross section, which can be used to constrain the non-perturbative Sudakov factor. 
Since one of the essential parts of the Sudakov factor is universal and spin-independent, 
its reliable extraction from spin-averaged cross sections will result in  
an improved  analysis of the Sivers asymmetries in the transverse spin-dependent scatterings. 
With this new Sudakov factor, we then perform a global fitting of the HERMES, COMPASS and JLab 
experimental data on polarized reactions to extract the Sivers functions. 
Finally, we reverse the sign of the quark Sivers functions to make predictions for the DY dilepton and $W$ 
boson production that will be measured in the near future to test the sign change of the Sivers effect.

\section{QCD evolution of TMDs: unpolarized differential cross sections}
In this section, we review the QCD evolution of TMDs \cite{collins-book,Aybat:2011zv,Aybat:2011ge,Echevarria:2012pw}.  We propose a simple non-perturbative 
Sudakov factor in the evolution formalism and demonstrate that it leads to a reasonably good description 
of the transverse momentum distribution for hadron production in SIDIS, as well as DY dilepton 
and $W/Z$ boson production in $pp$ collisions. We present a detailed comparison of our results 
with the experimental data on the hadron multiplicity distributions in SIDIS from both HERMES 
and COMPASS experiments, DY dilepton production at Fermilab fixed-target experiments and 
$W/Z$ boson production at the Tevatron and LHC energies. 

\subsection{QCD evolution of TMDs}
Our main focus is the transverse momentum dependent distribution function $F(x, k_\perp; Q)$ \cite{collins-book,Echevarria:2012js} \footnote{The properly defined TMDs depend on two scales \cite{collins-book,Aybat:2011zv,Aybat:2011ge,Echevarria:2012pw}, i.e., the factorization scale $\mu$ and another scale $\zeta$ related to the relevant high scale in the considered process, say the virtuality $Q$ of the photon in SIDIS process. We set them equal for simplicity, $\mu=\sqrt{\zeta}=Q$.}, 
which is probed at a momentum scale $Q$ and carries the collinear momentum fraction $x$ and 
a transverse component $k_\perp$. Since the evolution formalism is simpler in the coordinate space, 
we define the Fourier transform of $F(x, k_\perp; Q)$ in the two-dimensional coordinate space 
(referred to as $b$-space below) as
\bea
F(x, b; Q) = \int d^2k_\perp e^{-ik_\perp\cdot b}F(x, k_\perp; Q). 
\eea
The energy evolution of the TMD $F(x, b; Q)$ in the $b$-space has been derived by various 
groups and has the following form \cite{collins-book,Aybat:2011zv,Aybat:2011ge,Echevarria:2012pw}:
\bea
F(x, b; Q_f) = F(x, b; Q_i) \exp\left\{-\int_{Q_i}^{Q_f} \frac{d\mu}{\mu} 
\left(\Gamma_{\rm cusp}\ln\frac{Q_f^2}{\mu^2}+\gamma^V\right)\right\}
\left(\frac{Q_f^2}{Q_i^2}\right)^{-D(b;Q_i)}
\,,\quad\quad
\frac{dD}{d\ln\mu} = \Gamma_{\rm cusp}
\,,
\label{evolution}
\eea
Here $\Gamma_{\rm cusp}$ stands for the well-known cusp anomalous dimension with non cusp $\gamma^V$ \cite{Echevarria:2012pw}, $Q_i$ and $Q_f$ are the initial and final momentum scales for the QCD evolution, respectively.
It is important to emphasize that the evolution kernel in the right hand side of Eq.~\eqref{evolution} 
is valid only in the perturbative region, i.e., when $1/b\gg \Lambda_{\rm QCD}$. 

The function $F(x, b; Q)$ can represent any TMD. The relevant ones for this paper will be 
the unpolarized transverse momentum dependent PDFs and FFs, and the $k_\perp$-weighted Sivers 
function. They are defined as follows: 
\bea
f_{q/A}(x, b; Q) &= \int d^2k_\perp \, e^{-ik_\perp\cdot b} f_{q/A}(x, k_\perp^2; Q),
\label{pdfs}
\\
D_{h/q}(z, b; Q) &= \frac{1}{z^2} \int d^2p_T \, e^{-ip_T\cdot b/z} D_{h/q}(z, p_T^2; Q),
\label{ffs}
\\
f_{1T}^{\perp q(\alpha)}(x, b; Q) &= \frac{1}{M} \int d^2k_\perp \, e^{-ik_\perp\cdot b} 
k_{\perp}^\alpha f_{1T}^{\perp q}(x, k_\perp^2; Q),
\label{sivers}
\eea
where $f_{q/A}(x, k_\perp^2; Q)$ and $D_{h/q}(z, p_T^2; Q)$ are the unpolarized transverse momentum 
dependent PDF and FF in  momentum space, while $f_{1T}^{\perp q}(x, k_\perp^2; Q)$ is the quark Sivers 
function in the so-called Trento convention~\cite{Bacchetta:2004jz}. It is important to keep in 
mind that $f_{q/A}(x, b; Q)$, $D_{h/q}(z, b; Q)$, and $f_{1T}^{\perp q(\alpha)}(x, b; Q)$ follow 
exactly the same QCD evolution in the perturbative region as in Eq.~\eqref{evolution} \cite{Kang:2011mr,Aybat:2011ge,Echevarria:2012pw,evolutionpaper}.

In this paper we apply the well-known Collins-Soper-Sterman (CSS) approach~\cite{Collins:1984kg,Qiu:2000ga,Landry:2002ix} and choose an initial scale $Q_i = c/b$ to start the evolution of the TMDs. 
Here $c=2e^{-\gamma_E}$, with $\gamma_E \approx 0.577$ the Euler's constant. 
Thus, the evolution of a TMD from an initial scale $Q_i = c/b$ up to the scale $Q_f=Q$ is given by
\bea
\label{evolutionCSS}
F(x, b; Q) = F(x, b; c/b) \exp\left\{-\int_{c/b}^Q \frac{d\mu}{\mu} \left(A\ln\frac{Q^2}{\mu^2}+B\right)\right\}
\left(\frac{Q^2}{(c/b)^2}\right)^{-D(b;c/b)}
\,,
\eea
which we have written in terms of the conventional CSS notations with functions $A$ and $B$: $A=\Gamma_{\rm cusp}$ and $B=\gamma^V$ in Eq.~\eqref{evolution}. These functions, together with the $D$ term, are perturbatively expanded as in $A=\sum_{n=1}^\infty A^{(n)} \left(\alpha_s/\pi\right)^n$, $B=\sum_{n=1}^\infty B^{(n)} \left(\alpha_s/\pi\right)^n$ and
$D=\sum_{n=1}^\infty D^{(n)} \left(\alpha_s/\pi\right)^n$.
The coefficients we keep in our phenomenological analysis, which corresponds to  next-to-leading-logarithmic (NLL) accuracy~\cite{Echevarria:2012pw}, are given by~\cite{Kang:2011mr,Aybat:2011zv,Echevarria:2012pw,Collins:1984kg,Qiu:2000ga,Landry:2002ix}:
\bea
A^{(1)} &=C_F,
\\
A^{(2)} &=\frac{C_F}{2}\left[C_A\left(\frac{67}{18}-\frac{\pi^2}{6}\right)-\frac{10}{9}T_R n_f\right],
\\
B^{(1)} &= -\frac{3}{2} C_F,
\\
D^{(1)} &= \frac{C_F}{2}\ln\frac{Q_i^2 b^2}{c^2}
\quad \longrightarrow \quad
D^{(1)}(b; Q_i=c/b)=0
\,. 
\eea

In the region where $1/b\gg \Lambda_{\rm QCD}$ we can expand the initial TMD $F(x, b; \mu=c/b)$ in terms of the corresponding collinear function as follows
\bea
F_{i/h}(x, b; \mu) = 
\sum_a \int_x^1 \frac{d\xi}{\xi} 
C_{i/a}\left(\frac{x}{\xi},b; \mu \right) f_{a/h}(\xi, \mu)
+{\cal O}(b\,\Lambda_{\rm QCD})
\,,
\eea
where $C_{i/a}(z,b; \mu) = \sum_{n=0}^\infty C_{i/a}^{(n)}(\alpha_s/\pi)^n$ is the perturbatively 
calculable coefficient function with the leading order (LO) result $C_{i/a}^{(0)}=\delta_{ia}\delta(1-z)$ \cite{Kang:2011mr,Collins:1984kg, Qiu:2000ga,Nadolsky:2000ky}. 
Consistently with the NLL accuracy, in our phenomenological studies we only keep the LO results for the coefficient functions. In other words,
\bea
f_{q/A}(x, b; \mu) &=  f_{q/A}(x, \mu) + \cdots,
\label{pdf-expand}
\\
D_{h/q}(z, b; \mu) &= \frac{1}{z^2} D_{h/q}(z, \mu) + \cdots,
\label{ff-expand}
\\
f_{1T, \rm SIDIS}^{\perp q(\alpha)}(x, b; \mu)&= \left(\frac{ib^\alpha}{2}\right) T_{q,F}(x, x, \mu) + \cdots,
\label{sivers-expand}
\eea
where $``\cdots"$ represents the  contributions from higher order coefficients $C_{i/a}^{(n)}$ 
with $n\geq 1$ that are neglected in our current study. 
The functions $f_{q/A}(x, \mu)$ and $D_{h/q}(z, \mu)$ are the collinear PDFs and FFs, 
while $T_{q,F}(x, x, \mu)$ is the twist-3 Qiu-Sterman quark-gluon correlation function. 
Eq.~\eqref{sivers-expand} was first derived in~\cite{Kang:2011mr} and the result is 
not surprising because the Qiu-Sterman function is the first $k_\perp$-moment of the quark 
Sivers function~\cite{Boer:2003cm,Kang:2011hk}.
The subscript ``SIDIS'' on the left-hand side  emphasizes that the equation  is valid for 
the quark Sivers function measured in the SIDIS process. For the DY process there is an extra minus 
sign on the right-hand side.

With this in mind, we obtain the {\it perturbative} part  
(i.e., valid only when $1/b\gg \Lambda_{\rm QCD}$) of the TMD $F(x, b; Q)$ at NLL as 
\bea
F_{\rm pert}(x, b; Q) = f(x, c/b) \exp\left\{-\int_{c/b}^Q \frac{d\mu}{\mu} 
\left(A\ln\frac{Q^2}{\mu^2}+B\right)\right\},
\label{pert-evolution}
\eea
where $f(x, c/b)$ is the corresponding collinear function at scale $\mu=c/b$. 
In order to Fourier transform back and obtain the corresponding TMD $F(x, k_\perp; Q)$ in 
transverse momentum space,
\bea
F(x, k_\perp; Q) = \int\frac{d^2b}{(2\pi)^2} e^{ik_\perp\cdot b} F(x, b; Q) 
= \frac{1}{2\pi}\int_0^{\infty} db\, b J_0(k_\perp b) F(x, b; Q), 
\eea
with $J_0$ being the Bessel function of the zeroth order, one needs the information for the whole $b\in [0, \infty]$ region. Thus, to perform the 
Fourier transform, we have to extrapolate to the non-perturbative large-$b$ region. For this part, 
we follow the standard CSS approach~\cite{Collins:1984kg,Landry:2002ix} and introduce 
a non-perturbative Sudakov factor $R_{NP}(x, b, Q)$ as follows
\bea
F(x, b; Q) = F_{\rm pert}(x, b_{*}; Q) R_{NP}(x, b, Q) , 
\label{full-evolution}
\eea
where $b_{*} = b/\sqrt{1+(b/b_{\rm max})^2}$ and $b_{\rm max}$ is introduced such that 
$b_*\approx b$ at small $b\ll b_{\rm max}$ region, while it approaches the limit $b_{\rm max}$ 
when $b$ becomes non-perturbatively large. 
The value of $b_{\rm max}$ is typically chosen to be of order $\sim 1$ GeV$^{-1}$ and should 
be thought of as characterizing the boundary of the perturbative region of the $b$-dependence. 
The non-perturbative Sudakov factor $R_{\rm NP}(b, Q) = \exp(-S_{\rm NP})$ has been extensively 
studied. It has been extracted from the experimental data, in particular from $W/Z$ boson production 
at high energies~\cite{Landry:2002ix,Konychev:2005iy}, and  is mainly constrained by the large $Q$ 
fits. In this work we want to find a universal form, such that it can be used to describe the 
experimental data for SIDIS at relatively low $Q$, DY dilepton production at intermediate $Q$, 
and $W/Z$ boson production at large $Q$. A simple widely used non-perturbative Sudakov 
exponent $S_{\rm NP}$ has the following form~\cite{Landry:2002ix,Konychev:2005iy,Davies:1984sp,Ellis:1997sc}
\bea
S_{\rm NP}^{\rm pdf}(b, Q) &= b^2\left(g_1^{\rm pdf}+ \frac{g_2}{2} \ln\frac{Q}{Q_0}\right),
\\
S_{\rm NP}^{\rm ff}(b, Q) &= b^2\left(g_1^{\rm ff}+ \frac{g_2}{2} \ln\frac{Q}{Q_0}\right),
\\
S_{\rm NP}^{\rm sivers}(b, Q) &=b^2\left(g_1^{\rm sivers} + \frac{g_2}{2} \ln\frac{Q}{Q_0}\right),
\eea
for the unpolarized TMDPDFs, TMDFFs, and the weighted quark Sivers function as in 
Eqs.~\eqref{pdfs}, \eqref{ffs}, and \eqref{sivers}, respectively. Combining the $b_*$ prescription 
 with Eqs.~\eqref{pert-evolution} and \eqref{full-evolution}, we can write out the evolved 
TMDs explicitly as
\bea
f_{q/A}(x, b; Q) &= f_{q/A}(x, c/b_*) \exp\left\{-\int_{c/b_*}^Q \frac{d\mu}{\mu} 
\left(A\ln\frac{Q^2}{\mu^2}+B\right)\right\}
\exp\left\{-b^2\left(g_1^{\rm pdf}+ \frac{g_2}{2} \ln\frac{Q}{Q_0}\right)\right\},
\label{pdf-form}
\\
D_{h/q}(z, b; Q) &= \frac{1}{z^2}D_{h/q}(x, c/b_*) \exp\left\{-\int_{c/b_*}^Q \frac{d\mu}{\mu} 
\left(A\ln\frac{Q^2}{\mu^2}+B\right)\right\}
\exp\left\{-b^2\left(g_1^{\rm ff}+ \frac{g_2}{2} \ln\frac{Q}{Q_0}\right)\right\},
\label{ff-form}
\\
f_{1T, \rm SIDIS}^{\perp q(\alpha)}(x, b; Q) &= \left(\frac{ib^\alpha}{2}\right) T_{q,F}(x, x, c/b_*)
 \exp\left\{-\int_{c/b_*}^Q \frac{d\mu}{\mu} \left(A\ln\frac{Q^2}{\mu^2}+B\right)\right\}
\exp\left\{-b^2\left(g_1^{\rm sivers}+ \frac{g_2}{2} \ln\frac{Q}{Q_0}\right)\right\}.
\label{sivers-form}
\eea

It is important to realize that $g_2$ is universal for  all different types of TMDs
 and is certainly spin-independent, which is one of the important predictions of  QCD 
factorization theorems involving TMDs~\cite{collins-book, Aybat:2011zv}. On the other hand, 
the constant term $g_1$ depends on the type of TMDs, and can be interpreted as the intrinsic 
transverse momentum width for the relevant TMDs at the momentum scale 
$Q_0$~\cite{Qiu:2000ga,Aybat:2011zv,Anselmino:2012aa}. Assuming a Gaussian form, we have
\bea
g_1^{\rm pdf} = \frac{\langle k_\perp^2\rangle_{Q_0}}{4},
\qquad
g_1^{\rm ff} = \frac{\langle p_T^2\rangle_{Q_0}}{4z^2},
\qquad
g_1^{\rm sivers} = \frac{\langle k_{s\perp}^2\rangle_{Q_0}}{4},
\eea
where $\langle k_\perp^2\rangle_{Q_0}$, $\langle p_T^2\rangle_{Q_0}$, and $\langle k_{s\perp}^2\rangle_{Q_0}$ 
are the relevant averaged intrinsic transverse momenta squared for TMDPDFs, TMDFFs, and the quark Sivers functions at the momentum scale $Q_0$, respectively. 

Once we resort to such an intuitive interpretation and further choose $Q_0 = \sqrt{2.4}$ GeV, 
the typical virtuality scale in the HERMES experiments,  $\langle k_\perp^2\rangle_{Q_0}$ and 
$\langle p_T^2\rangle_{Q_0}$ have been extracted from the HERMES experimental data by various 
groups~\cite{Anselmino:2005nn, Collins:2005ie, Schweitzer:2010tt, note-new}. At present,  values in the 
following ranges can give an equally good description of the data: 
\bea
\langle k_\perp^2\rangle_{Q_0} = 0.25 - 0.44 {\rm ~GeV}^2,
\qquad
\langle p_T^2\rangle_{Q_0} = 0.16 - 0.20 {\rm ~GeV}^2.
\label{width}
\eea
On the other hand, the universal parameter $g_2$ has been extracted mainly from the DY lepton pair 
and $W/Z$ production. The value of $g_2$ is intimately connected to the value $b_{\rm max}$ one is using. 
In Ref.~\cite{Konychev:2005iy}, Konychev and Nadolsky have shown that the best fit of the experimental 
data can be reached if one chooses $b_{\rm max}=1.5$ GeV$^{-1}$, and the fitted $g_2$ is given by
\bea
g_2 = 0.184\pm 0.018 {\rm ~GeV}^2 .
\label{g2}
\eea

In our work, we will try to tune the three parameters 
$\langle k_\perp^2\rangle_{Q_0}$, $\langle p_T^2\rangle_{Q_0}$, and $g_2$ within their current 
extracted ranges, Eqs.~\eqref{width} and \eqref{g2}, to see if we can indeed reconcile 
the SIDIS process and the DY-type processes, and to test if we can describe all the SIDIS, DY 
lepton pair, and $W/Z$ production data. Indeed, we find the following parameters can do a 
rather reasonable job:
\bea
\langle k_\perp^2\rangle_{Q_0} = 0.38 {\rm ~GeV}^2,
\qquad
\langle p_T^2\rangle_{Q_0} = 0.19 {\rm ~GeV}^2,
\qquad
g_2 = 0.16  {\rm ~GeV}^2.
\label{sudakov}
\eea
The variation of the non-perturbative parameters that enter into the evolution of TMDs should not affect the shape of the kernel in the perturbative region $1/b\gg \Lambda_{\rm QCD}$, where no non-perturbative model is needed. 
In other words, the relative change of the parameters $b_{\rm max}$ and $g_2$ should conspire in such a way that the kernel in the perturbative region is not spoiled. 
We have checked this fact explicitly at NLL accuracy for our tuned parameters $b_{\rm max}=1.5~{\rm GeV}^{-1}$ and $g_2=0.16$, and found that this is indeed the case. 
In the next subsection we show that the implementation of the Sudakov factor with the above $g_2$ parameter leads to a reasonably good description of all the experimental data on SIDIS, DY lepton pair and $W/Z$ boson production, and hence a more solid extraction of the Sivers asymmetry.

\subsection{Transverse momentum distribution}
Here we first review the QCD factorization formalism for the transverse momentum distribution 
of hadron production in SIDIS, DY lepton pair and $W/Z$ boson production in $pp$ collisions. We then 
demonstrate that the QCD factorization formalism with the evolution implemented as in 
Eq.~\eqref{full-evolution} and the tuned non-perturbative Sudakov factor with parameters 
given in Eq.~\eqref{sudakov} leads to a reasonably good description of the experimental data on SIDIS, 
DY lepton pair, and $W/Z$ production. 

We start with single hadron production in SIDIS: the scattering processes of a lepton $e$ on a hadron $A$, 
\bea
e(\ell) + A(P)\to e(\ell') + h(P_h) +X, 
\eea
where we use $A$ (also $B$ below) generically to represent the incoming hadrons, and $h$ is the observed hadron with momentum $P_h$. We define the virtual photon momentum 
$q=\ell - \ell'$ and its invariant mass $Q^2 = - q^2$, and adopt the usual SIDIS variables \cite{Meng:1991da}:
\bea
S_{ep} = (P+\ell)^2,
\qquad
x_B = \frac{Q^2}{2P\cdot q},
\qquad
y=\frac{P\cdot q}{P\cdot \ell} = \frac{Q^2}{x_B S_{ep}},
\qquad
z_h = \frac{P\cdot P_h}{P\cdot q}.
\eea
The so-called hadron multiplicity distribution is defined as
\bea
\frac{dN}{dz_h d^2P_{h\perp}} = \left. \frac{d\sigma}{dx_B dQ^2 dz_h d^2P_{h\perp}}\right/ 
\frac{d\sigma}{dx_B dQ^2},
\label{hadron-mul}
\eea
where the numerator and denominator are given by
\bea
\frac{d\sigma}{dx_B dQ^2 dz_h d^2P_{h\perp}} &= \frac{\sigma_0^{\rm DIS}}{2\pi} 
\sum_q e_q^2 \int_0^{\infty} db\, b J_0(P_{h\perp}b/z_h) f_{q/A}(x_B, b; Q) D_{h/q}(z_h, b; Q),
\label{sidis-pt}
\\
\frac{d\sigma}{dx_B dQ^2} &= \sigma_0^{\rm DIS} \sum_q e_q^2 f_{q/A}(x_B, Q),
\eea
with $\sigma_0^{\rm DIS} = 2\pi\alpha_{\rm em}^2\left[1+(1-y)^2\right]/Q^4$. Here, $f_{q/A}(x_B, Q)$ is the collinear PDF at momentum scale $Q$, 
while $f_{q/A}(x_B, b; Q)$ and $D_{h/q}(z_h, b; Q)$ are the evolved TMDPDFs and TMDFFs given by 
Eqs.~\eqref{pdf-form} and \eqref{ff-form}, respectively. It is worth pointing out that we have taken the hard factor at LO (equal to 1) in the above TMD factorization formalism as in Eq.~\eqref{sidis-pt} and throughout the paper, to be consistent with the fact that we use the LO coefficient function in Eqs.~\eqref{pdf-expand} - \eqref{sivers-expand}. Notice as well that the relevant soft function for each process, either DY or SIDIS, which accounts for the soft gluon radiation, is already included in the proper definition of the TMDs in each case \cite{collins-book,Echevarria:2012js}.

On the other hand, for Drell-Yan lepton pair production, 
$A(P_A)+B(P_B)\to [\gamma^*\to] \ell^+\ell^-(y,Q, p_\perp) +X$, with $y, Q, p_\perp$ being the rapidity, 
invariant mass and transverse momentum of the pair, respectively, the spin-averaged 
differential cross section can be written as \cite{GarciaEchevarria:2011rb}
\bea
\frac{d\sigma}{dQ^2 dy d^2p_\perp} = \frac{\sigma_0^{\rm DY}}{2\pi}
\sum_q e_q^2 \int_0^{\infty} db\, bJ_0(p_\perp b) f_{q/A}(x_a, b; Q) f_{\bar q/B}(x_b, b; Q).
\label{DY}
\eea
Here $\sigma_0^{\rm DY} = 4\pi\alpha_{\rm em}^2/3sQ^2N_c$, $s=(P_A+P_B)^2$ is the center-of-mass 
(CM) energy squared, and the parton momentum fractions $x_a$ and $x_b$ are given by
\bea
x_a = \frac{Q}{\sqrt{s}} e^y,
\qquad
x_b = \frac{Q}{\sqrt{s}} e^{-y}.
\label{xab}
\eea
Likewise, $f_{q/A}(x_a, b; Q)$ and $f_{\bar q/B}(x_b, b; Q)$ are the QCD evolved TMDPDFs in 
Eq.~\eqref{pdf-form}. Similarly, for $W/Z$ production, $A(P_A)+B(P_B)\to W/Z(y, p_\perp)+X$, 
the differential cross sections are given by~\cite{Kang:2009bp,Kang:2009sm}
\bea
\frac{d\sigma^W}{dyd^2p_\perp} &= \frac{\sigma_0^W}{2\pi} \sum_{q,q'} 
|V_{qq'}|^2 \int_0^{\infty} db\, b J_0(q_\perp b) f_{q/A}(x_a, b; Q) f_{q'/B}(x_b, b; Q),
\label{w-cross}
\\
\frac{d\sigma^Z}{dyd^2p_\perp} &= \frac{\sigma_0^Z}{2\pi} \sum_{q} \left(V_q^2+A_q^2\right) 
\int_0^{\infty} db\, b J_0(q_\perp b) f_{q/A}(x_a, b; Q) f_{\bar q/B}(x_b, b; Q),
\label{z-cross}
\eea
where $V_{qq'}$ are the CKM matrix elements for the weak interaction, and $V_q$ and $A_q$ are 
the vector and axial couplings of the $Z$ boson to the quark, respectively. 
The LO cross sections $\sigma_0^W$ and $\sigma_0^Z$ have the following form
\bea
\sigma_0^W = \frac{\sqrt{2} \pi G_F M_W^2}{sN_c},
\qquad
\sigma_0^Z = \frac{\sqrt{2} \pi G_F M_Z^2}{sN_c},
\eea
where $G_F$ is the Fermi weak coupling constant, and $M_W$ ($M_Z$) is the mass of the $W$ ($Z$) boson.

\bef
\psfig{file=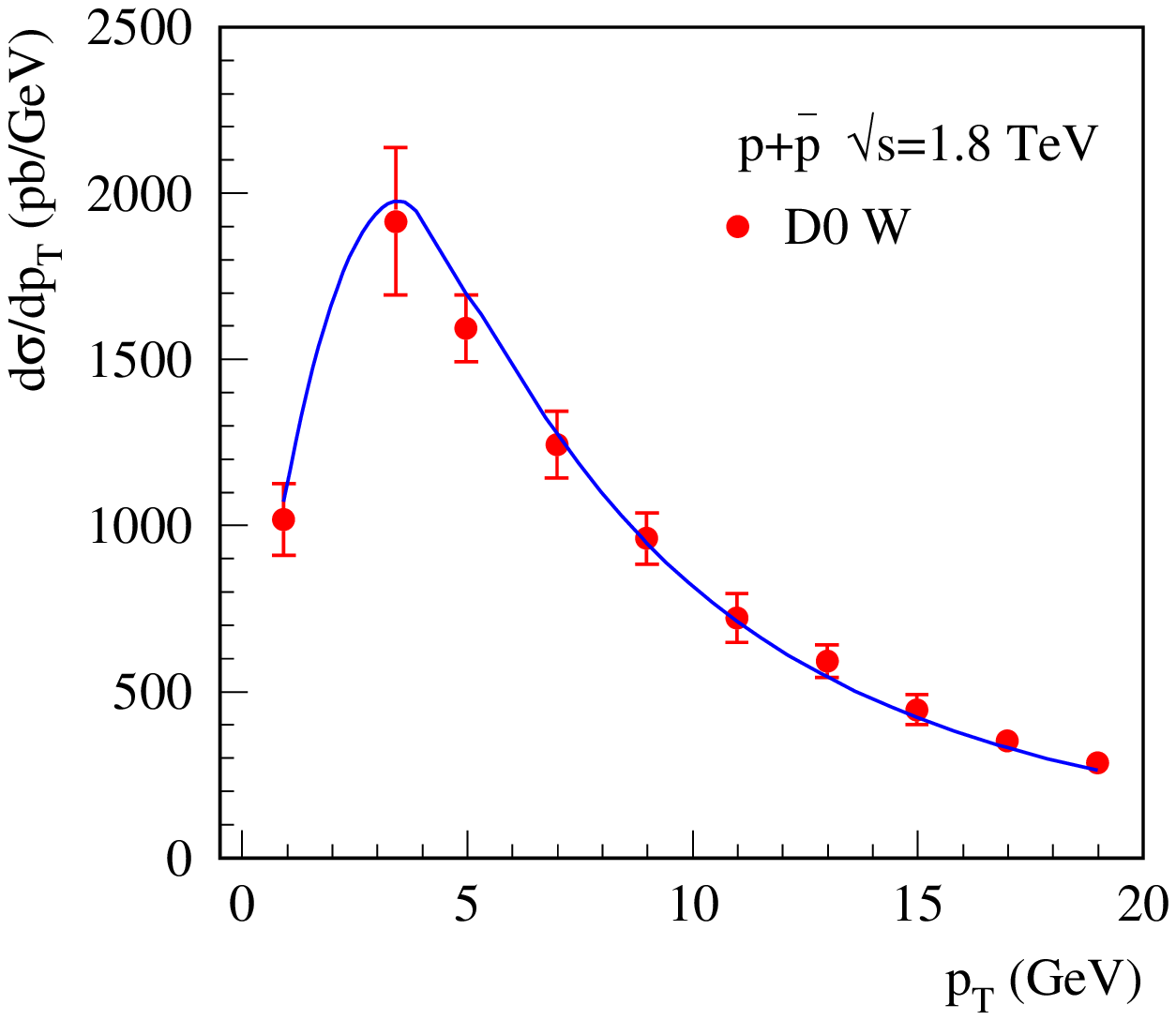, width=2.25in}
\hskip 0.1in
\psfig{file=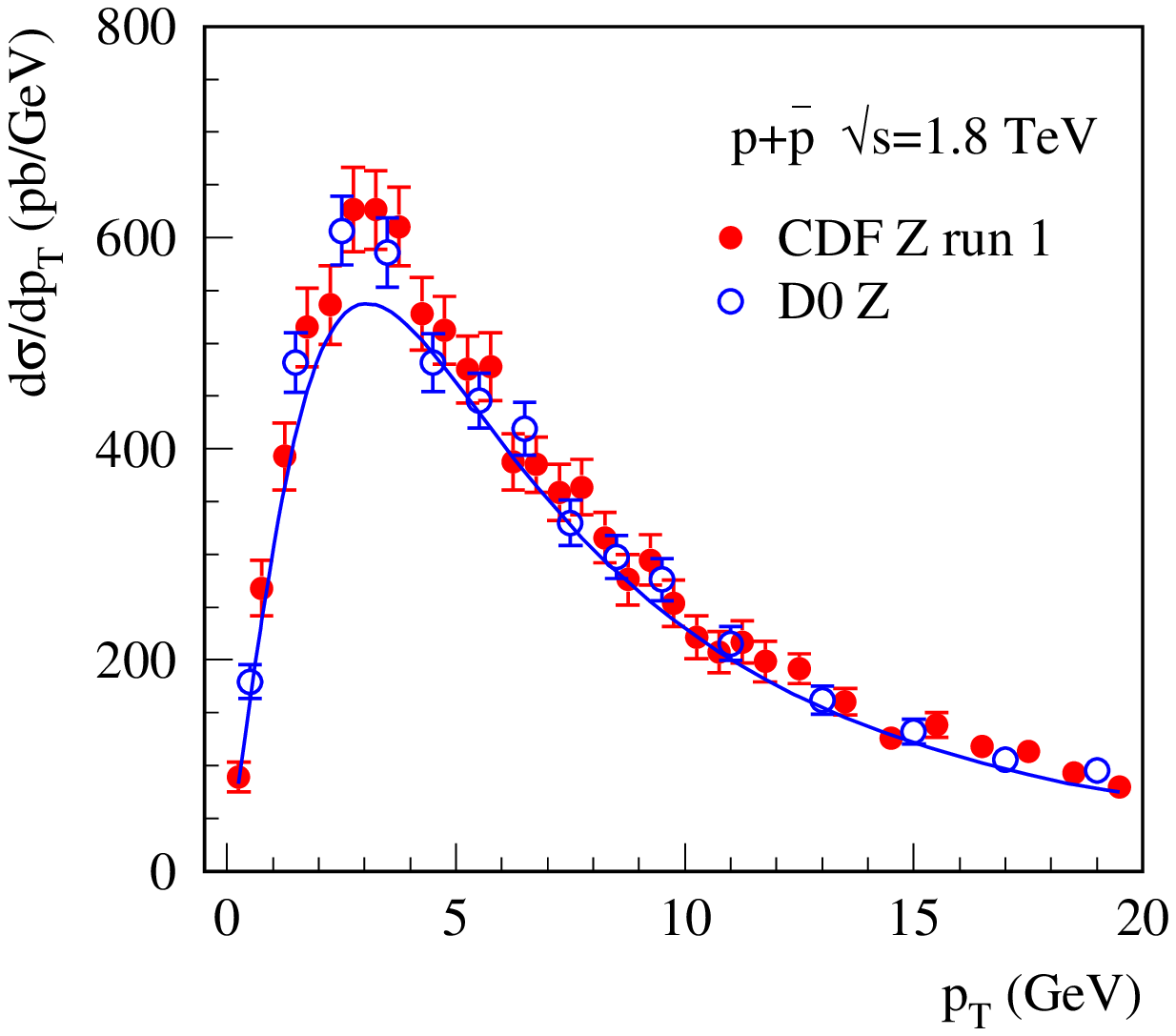, width=2.2in}
\hskip 0.1in
\psfig{file=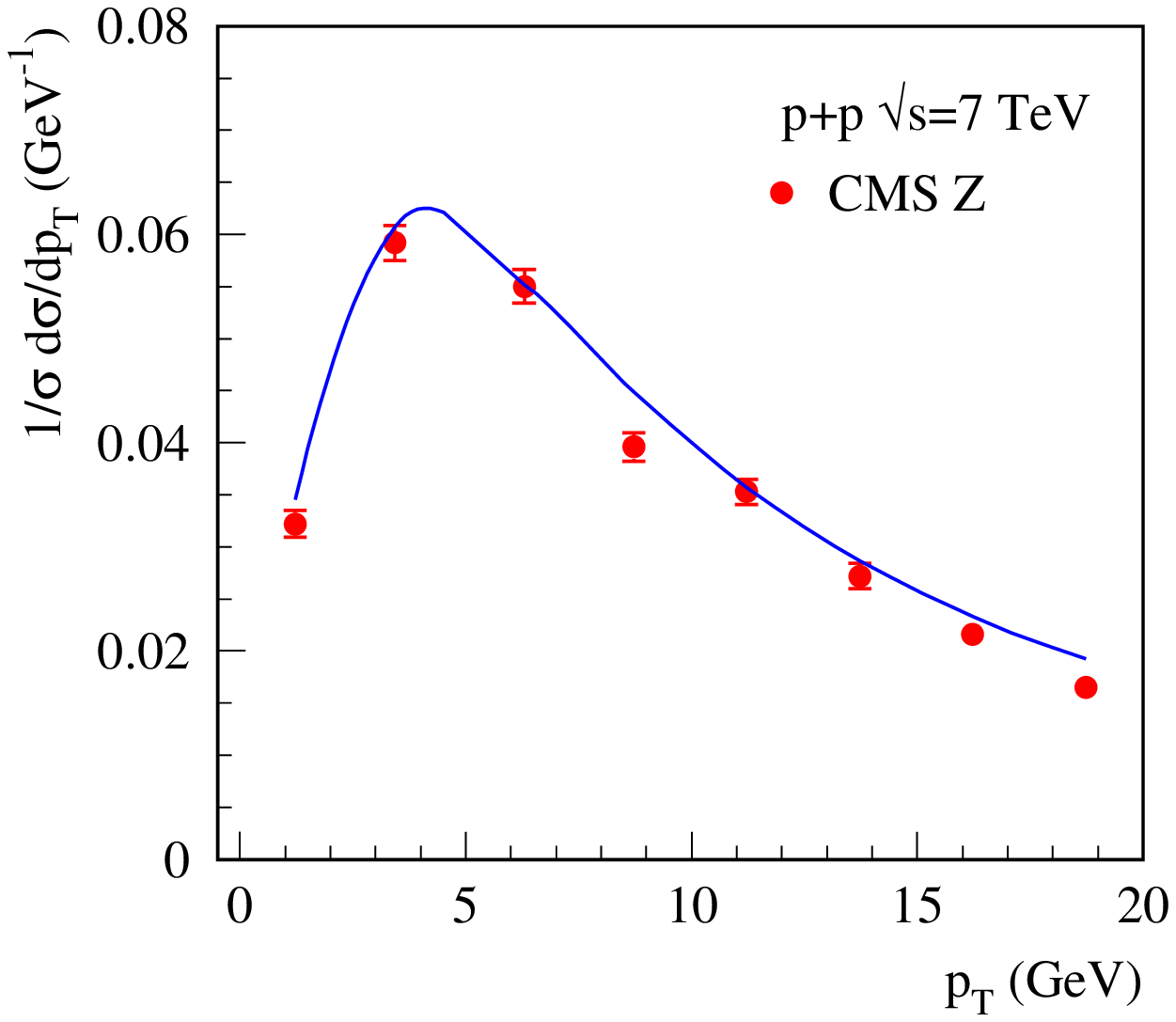, width=2.2in}
\caption{Comparison of theoretical results to $W$~\cite{Abbott:2000xv} (left) and 
$Z$~\cite{Abbott:1999wk,Affolder:1999jh} (middle) production in $p+\bar{p}$ collisions at $\sqrt{s}=1.8$ TeV, 
and $Z$ production~\cite{Chatrchyan:2011wt} (right) in $p+p$ collisions at $\sqrt{s} = 7$ TeV.}
\label{fig-WZ}
\eef
To compare with experimental data, we use the unpolarized parton 
distribution functions $f_{q/A}(x, Q)$ as given by the MSTW2008 parametrization~\cite{Martin:2009iq} 
and the DSS unpolarized fragmentation functions $D_{h/q}(z, Q)$~\cite{deFlorian:2007aj}. 
It is important to remember that our QCD factorization formalism based on TMDs is only applicable in 
the kinematic region where $p_\perp\ll Q$ \cite{collins-book}. To describe the large $p_\perp\sim Q$ region, 
one needs the complete next-to-leading order calculation, more precisely the so-called 
$Y$-term~\cite{Collins:1984kg,Qiu:2000ga,Landry:2002ix,Kang:2012am}. 
To be consistent with our formalism, we thus restrict our comparison with the experimental data as follows: 
for $W/Z$ boson production, we choose $p_\perp \leq 20$ GeV; for DY dilepton production, we have 
$p_\perp \leq 1.3$ GeV; for hadron production at COMPASS with $\langle Q^2\rangle = 7.57$ GeV$^2$, 
we choose $P_{h\perp}\leq 0.7$ GeV; for hadron production at HERMES 
with $\langle Q^2\rangle = 2.45$ GeV$^2$, we choose $P_{h\perp}\leq 0.6$ GeV such that we still have 
enough experimental data for the analysis. 

\bef
\psfig{file=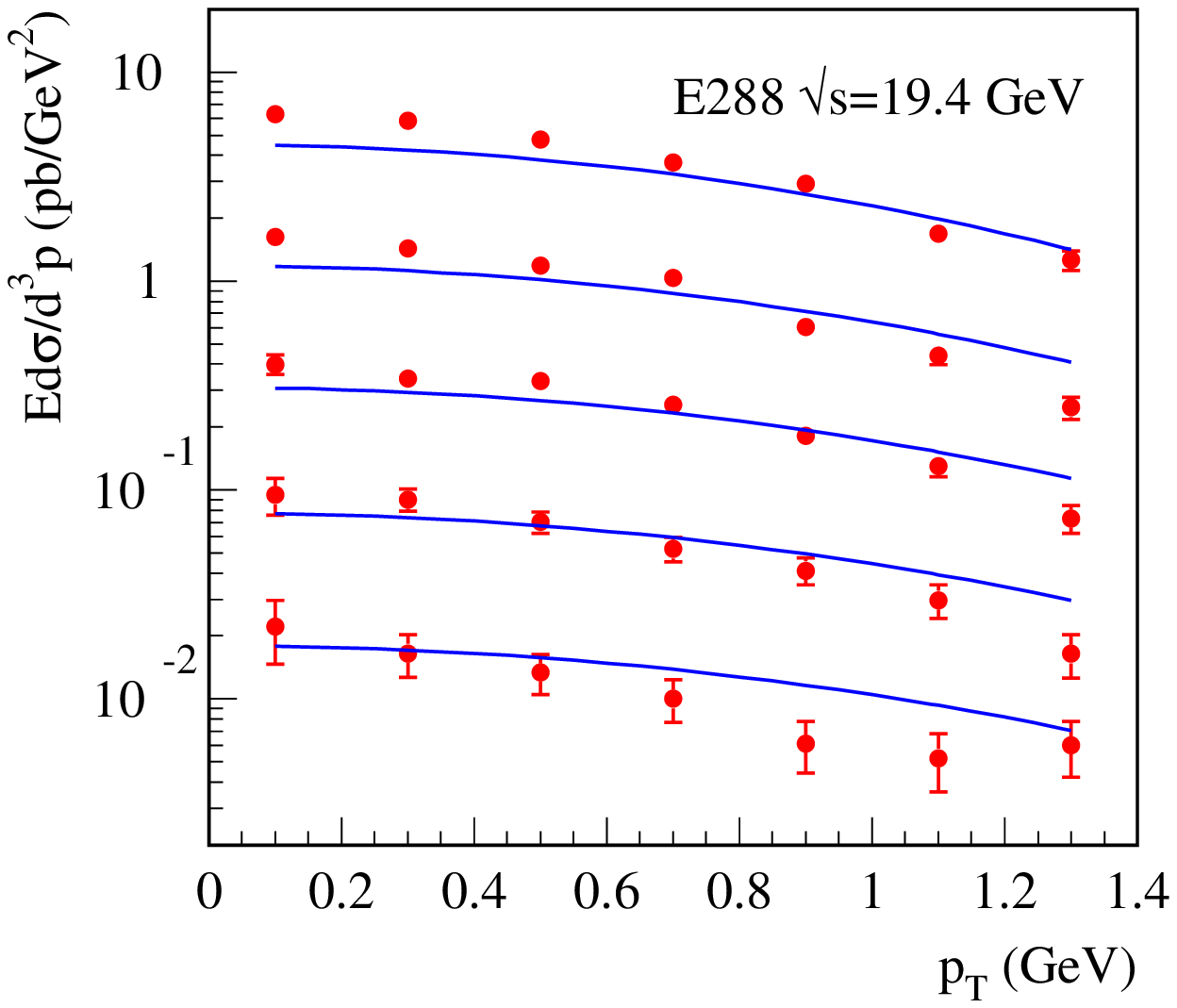, width=2.2in}
\hskip 0.2in
\psfig{file=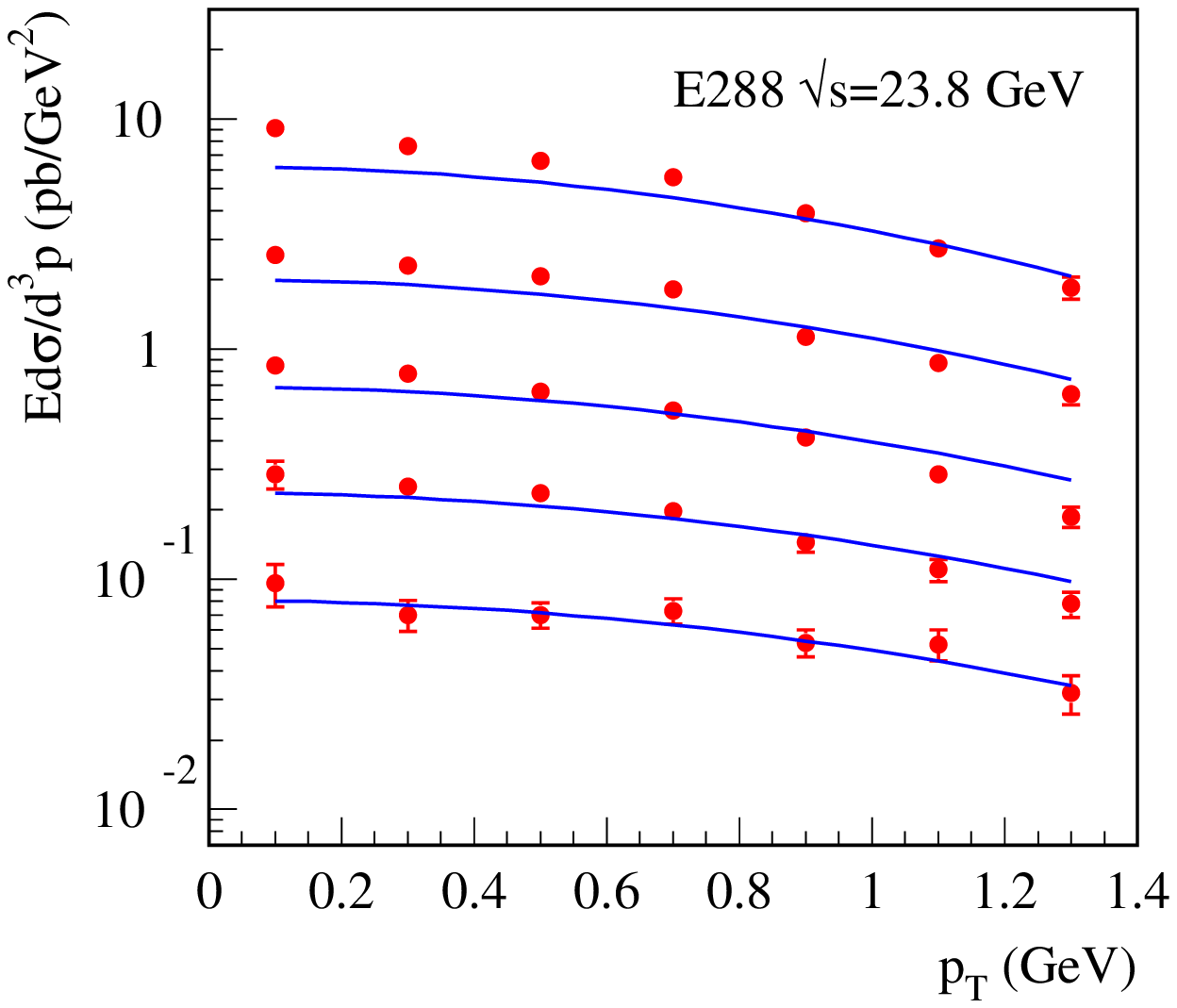, width=2.2in}
\vskip 0.1in
\psfig{file=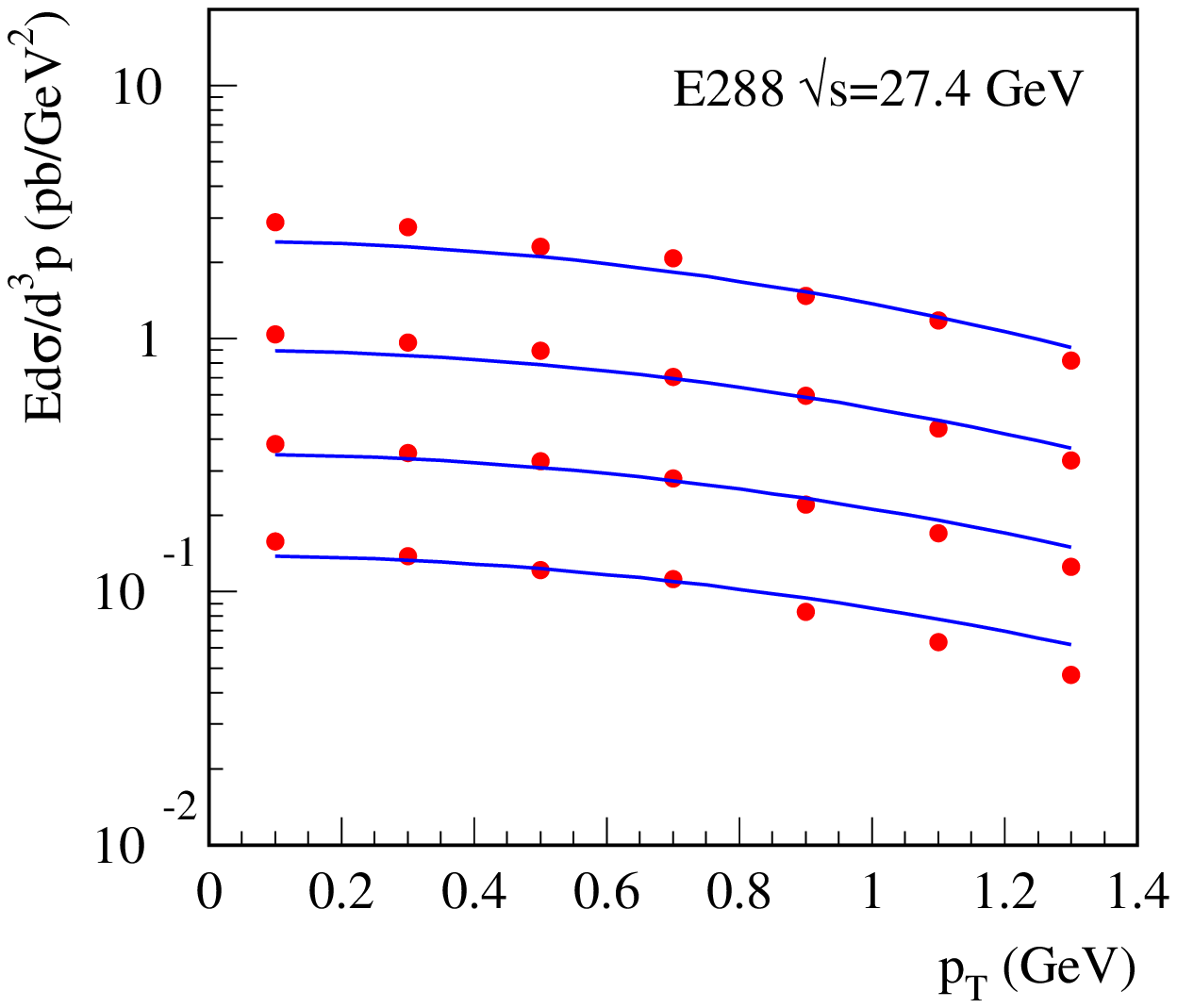, width=2.2in}
\hskip 0.2in
\psfig{file=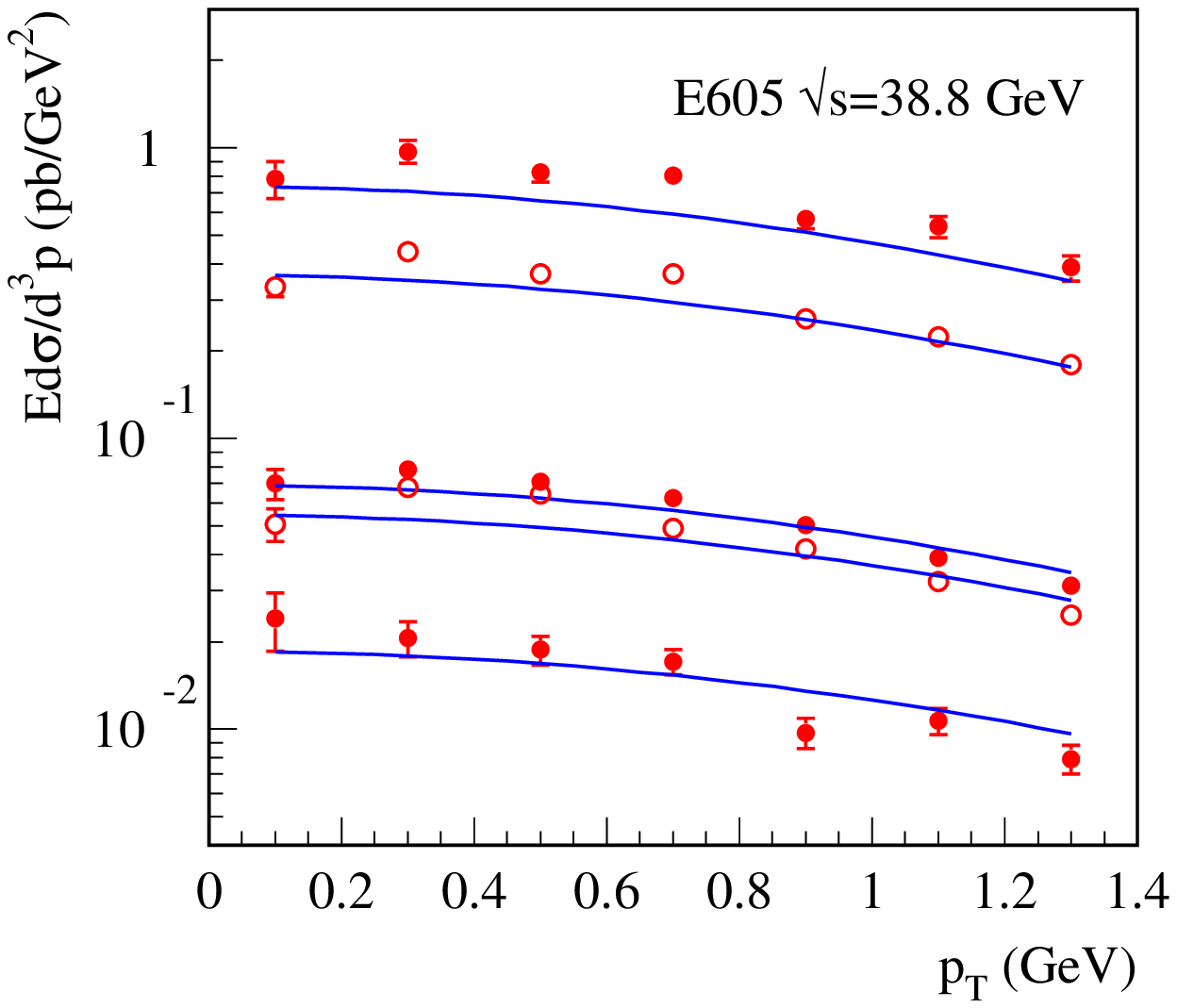, width=2.2in}
\caption{The first three plots show  comparisons with the Fermilab E288 Drell-Yan dilepton data at 
different CM energies $\sqrt{s}=19.4$ (left), 23.8, and 27.4 GeV~\cite{Ito:1980ev}. The data points 
from top to bottom correspond to different invariant mass $Q$ of the lepton pair. For the top two plots, 
they are: $[4, 5]$, $[5, 6]$, $[6, 7]$, $[7, 8]$, and $[8, 9]$ GeV. For the left bottom plot, 
it starts with the $[5, 6]$ GeV range (no $[4, 5]$ GeV range. The right bottom plot is the 
comparison with  the Fermilab E605 Drell-Yan dilepton data at CM energy $\sqrt{s}=38.8$ GeV~\cite{Moreno:1990sf}. 
Again the mass ranges are: $[7, 8]$, $[8, 9]$, $[10.5, 11.5]$, $[11.5, 13.5]$, and $[13.5, 18]$ GeV.}
\label{fig-DY}
\eef
We first compare in  Fig.~\ref{fig-WZ} our calculation, based on the QCD factorization 
formalism, Eqs.~\eqref{w-cross}  and \eqref{z-cross}, with  $W/Z$ production at both the Tevatron 
and LHC energies. With QCD evolved TMDPDFs given in Eq.~\eqref{pdf-form} and the tuned parameters 
for the Sudakov factor in Eq.~\eqref{sudakov}, we plot the $W$ and $Z$ boson differential cross section 
as a function of transverse momentum $p_\perp$. The left and middle panels of Fig.~\ref{fig-WZ}
are the comparisons with the $W/Z$ measurements~\cite{Abbott:2000xv,Abbott:1999wk,Affolder:1999jh} 
in $p+\bar p$ collisions at the Tevatron energy $\sqrt{s}=1.8$~TeV.  In the right 
panel of Fig.~\ref{fig-WZ} we compare with the most recent $Z$ boson measurement~\cite{Chatrchyan:2011wt} 
in $p+p$ collisions from the CMS collaboration at LHC energy $\sqrt{s}=7$~TeV. Our formalism gives a reasonably good description of the $W/Z$ boson production at both the Tevatron and LHC energies. 

Next, we  compare our calculation for the DY lepton pair production with the fixed-target Fermilab 
experimental data at different CM energies $\sqrt{s} = 19.4, 23.8, 27.4$ for the E288 
collaboration~\cite{Ito:1980ev} and at $\sqrt{s}=38.8$ GeV for the E605 collaboration~\cite{Moreno:1990sf}, 
see Fig.~\ref{fig-DY}. Since these experiments were really performed for $p+Cu$ collisions, 
we use the EKS98 parametrization~\cite{Eskola:1998df} for the collinear nuclear PDFs in the nucleus $Cu$. 
For both $\sqrt{s}=19.4$ and 23.8 GeV, the curves from top to bottom  correspond to the different invariant 
mass bins, i.e., $Q\in [4,5]$, $[5, 6]$, $[6, 7]$, $[7, 8]$, and $[8, 9]$ GeV. For $\sqrt{s}=27.4$ GeV, 
we have  $Q\in [5, 6]$, $[6, 7]$, $[7, 8]$, and $[8, 9]$ GeV. Finally, for $\sqrt{s}=38.8$ GeV the mass 
ranges are: $Q\in [7, 8]$, $[8, 9]$, $[10.5, 11.5]$, $[11.5, 13.5]$, and $[13.5, 18]$ GeV. As can be seen, 
our QCD formalism gives a reasonably good description of the Drell-Yan dilepton production in all the 
measured mass ranges. 

\bef
\psfig{file=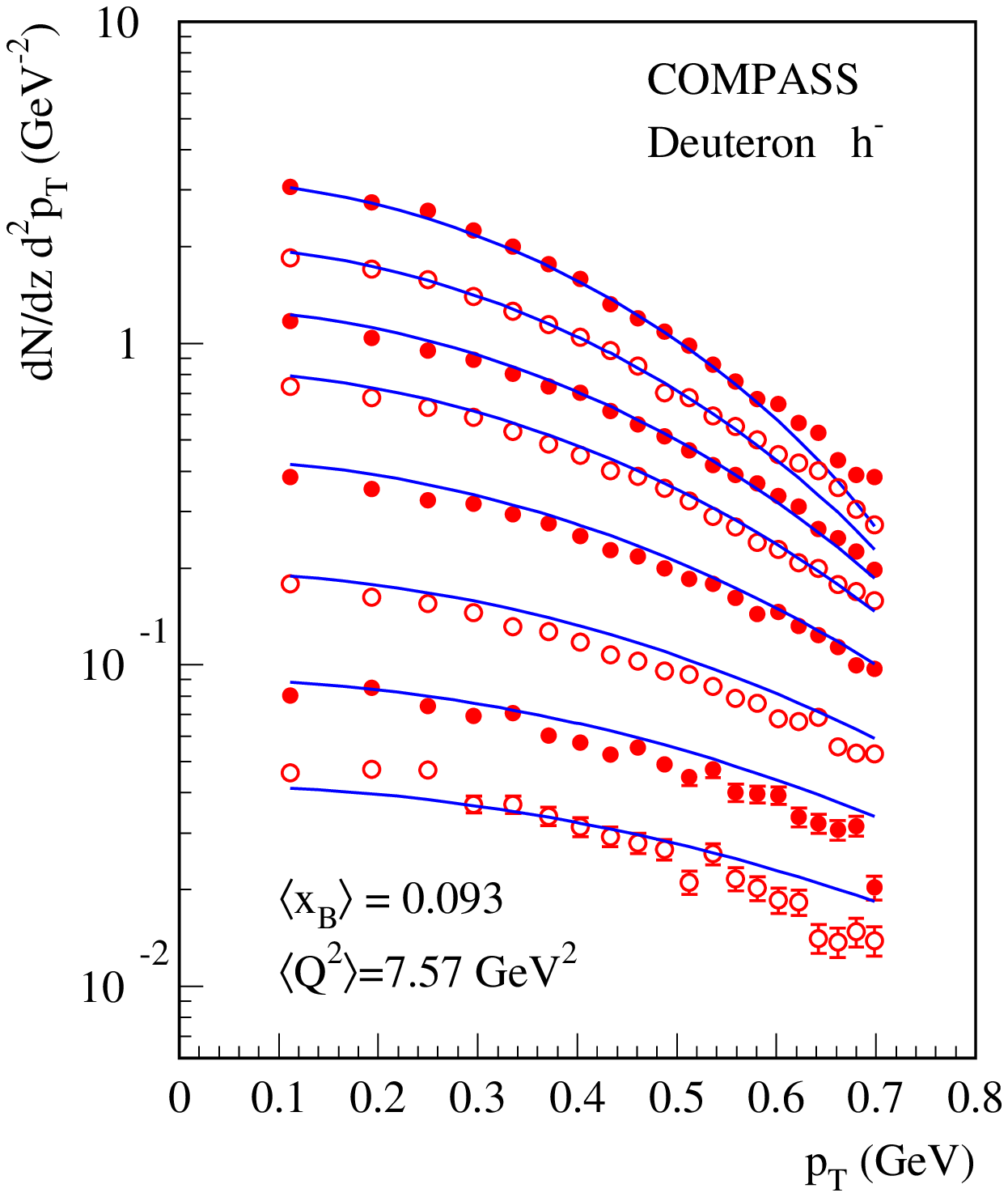, width=2.5in}
\hskip 0.2in
\psfig{file=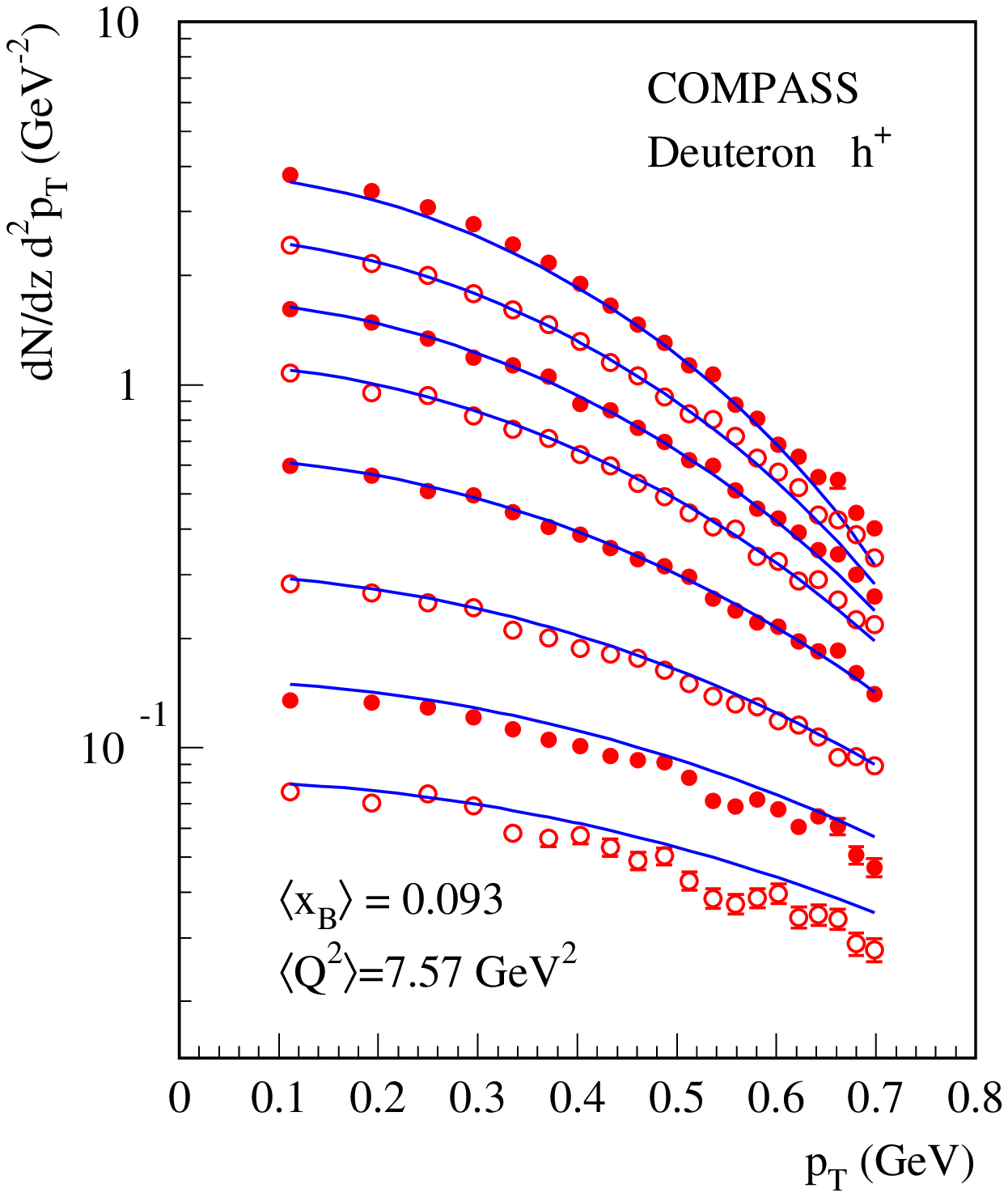, width=2.5in}
\caption{Comparison of the theoretical results with the COMPASS data (deuteron target)~\cite{Adolph:2013stb} 
at $\langle Q^2\rangle = 7.57$ GeV$^2$ and $\langle x_B\rangle = 0.093$. The data points from top to 
bottom correspond to different $z_h$ regions: $[0.2, 0.25]$, $[0.25, 0.3]$, $[0.3, 0.35]$, 
$[0.35, 0.4]$, $[0.4, 0.5]$, $[0.5, 0.6]$, $[0.6, 0.7]$, and $[0.7, 0.8]$.}
\label{fig-compass}
\eef
Let us now turn to the hadron multiplicity distribution in the SIDIS processes. In Fig.~\ref{fig-compass}, 
we compare our calculations with the recent COMPASS experimental data for the charged hadron multiplicity 
distribution~\cite{Adolph:2013stb} at $\langle Q^2\rangle = 7.57$ GeV$^2$ and $\langle x_B\rangle = 0.093$ 
for a deuteron target. The data points from top to bottom correspond to different $z_h$ regions: 
$z_h\in [0.2, 0.25]$, $[0.25, 0.3]$, $[0.3, 0.35]$, $[0.35, 0.4]$, $[0.4, 0.5]$, $[0.5, 0.6]$, $[0.6, 0.7]$, 
and $[0.7, 0.8]$. We find that for both negative and positive charged hadrons the QCD formalism 
in Eq.~\eqref{hadron-mul} gives a good description for the $P_{h\perp}$-dependence of the 
hadron multiplicity distribution. 

\bef
\psfig{file=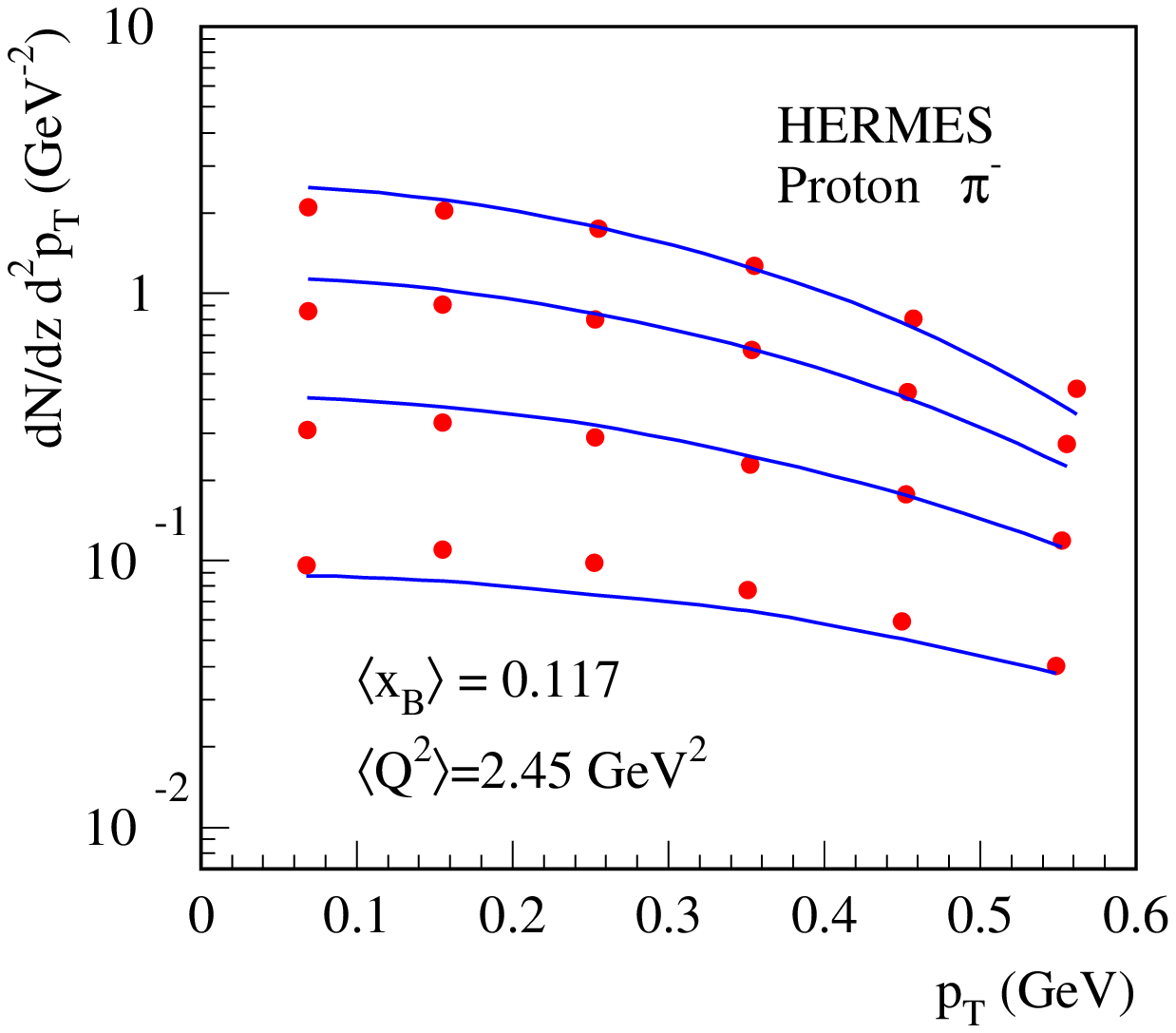, width=2.3in}
\hskip 0.2in
\psfig{file=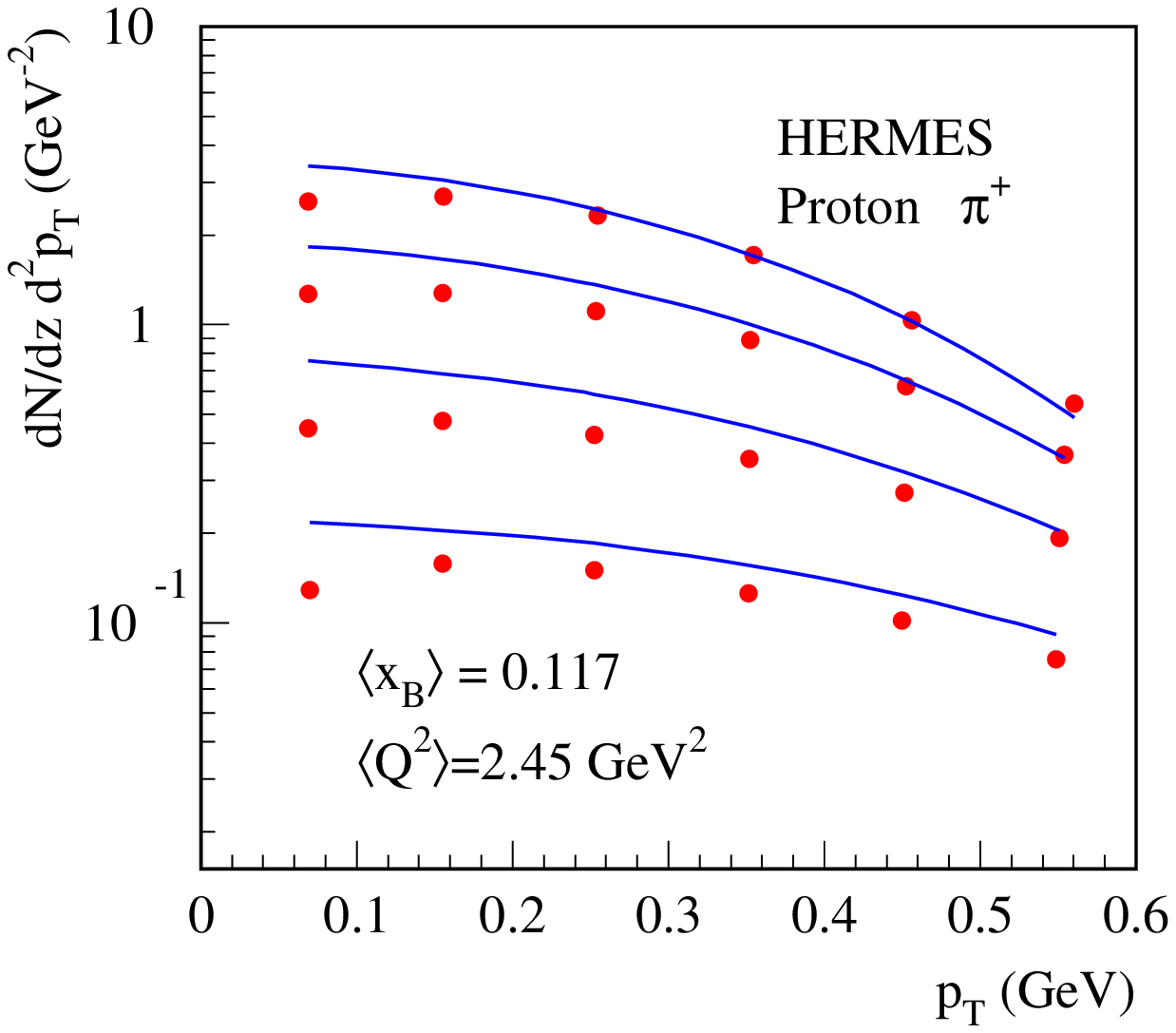, width=2.3in}
\caption{Comparison of theoretical results with the HERMES data (proton target)~\cite{Airapetian:2012ki} 
at $\langle Q^2\rangle = 2.45$ GeV$^2$ and $\langle x_B\rangle = 0.117$. The data points from top 
to bottom correspond to different $z_h$ regions: $[0.2, 0.3]$, $[0.3, 0.4]$, $[0.4, 0.6]$, and $[0.6, 0.8]$.}
\label{fig-hermes}
\eef
Finally, in Fig.~\ref{fig-hermes} we compare our calculation with the HERMES multiplicity distribution 
data \cite{Airapetian:2012ki} for a proton target at $\langle Q^2\rangle = 2.45$ GeV$^2$ and 
$\langle x_B \rangle =0.117$. The data points from top to bottom correspond to different $z_h$ regions: 
$z_h\in [0.2, 0.3]$, $[0.3, 0.4]$, $[0.4, 0.6]$, and $[0.6, 0.8]$. We find that our formalism 
still gives a reasonable description for $\pi^-$ multiplicity distribution data as a function of $P_{h\perp}$, 
though $\pi^+$ becomes worse when going to the high $z_h$ region. Note, however, that the normalization of such distributions is related to the fragmentation functions \cite{Airapetian:2012ki}. 

In summary we find that our proposed non-perturbative Sudakov factor in Eq.~\eqref{sudakov} along 
with $b_{\rm max}=1.5$ GeV$^{-1}$ gives a reasonably good description of the  hadron multiplicity 
distribution in SIDIS at rather low $Q$, DY lepton pair production at intermediate $Q$, and $W/Z$ production 
at high $Q$ from rather low CM energies up to the LHC energies. 
Even though the description is not perfect, 
one has to keep in mind that our QCD formalism is the very first attempt to use a universal form to 
describe  the experimental data on both SIDIS and DY-type processes. 
At the moment, we are implementing the evolution at NLL accuracy along with the LO coefficient functions. 
All of these could be further improved, and a first attempt to implement the approach presented in \cite{Echevarria:2012pw} is being pursued in \cite{madridtorino}.
Another important consequence is 
that since the parameter $g_2$ is a universal parameter, i.e. independent of the spin, we can then
 use the same $g_2$ to extract the Sivers functions from the current Sivers asymmetry measurements in SIDIS. 
This will be the main focus of the next section.

\section{QCD evolution of TMDs: the Sivers effect}
In this section we will first extract the quark Sivers functions from the Sivers asymmetry 
measurements in SIDIS from JLab, HERMES, and COMPASS experiments. We will then make predictions 
for the Sivers asymmetries of DY dilepton and $W$ boson production, to be compared with the future 
measurements. 

\subsection{Global fitting of Sivers asymmetries in SIDIS}
Here we apply our QCD evolution formalism for the Sivers effect in SIDIS 
and use it to extract the quark Sivers functions from the experimental data. 
The differential SIDIS cross section on a transversely polarized nucleon target can be written 
as~\cite{Kang:2012xf,Bacchetta:2006tn,Anselmino:2008sga}
\bea
\frac{d\sigma}{dx_B dy dz_h d^2P_{h\perp}}
&= \sigma_0(x_B, y, Q^2)
\left[F_{UU} +    \sin(\phi_h-\phi_s)\,
F_{UT}^{\sin\left(\phi_h -\phi_s\right)} \right],
  \label{eq:aut}
\eea
where $\sigma_0 = \frac{2\pi \alpha_{\rm em}^2}{Q^2 y}\left(1+(1-y)^2\right)$, 
and $\phi_s$ and $\phi_h$ are the azimuthal angles for the nucleon spin and the transverse momentum 
of the outgoing hadron, respectively.
$F_{UU}$ and $F_{UT}^{\sin(\phi_h-\phi_s)}$ are the spin-averaged and transverse spin-dependent 
structure functions that have the expressions:
\bea
F_{UU} &=  \frac{1}{2\pi} \int_0^\infty db\, b J_0(P_{h\perp} b/z_h)\sum_q e_q^2  
f_{q/A}(x_B, b; Q) D_{h/q}(z_h, b; Q),
\\
F_{UT}^{\sin\left(\phi_h -\phi_s\right)} & =  - \int\frac{d^2b}{(2\pi)^2} 
e^{iP_{h\perp}\cdot b/z_h} \hat{P}_{h\perp}^{\alpha}\sum_q e_q^2  
f_{1T, \rm SIDIS}^{\perp q(\alpha)}(x_B, b; Q) D_{h/q}(z_h, b; Q),
\label{fut}
\eea
where $\hat{P}_{h\perp}$ is the unit vector along the hadron transverse momentum $P_{h\perp}$. 
If we include the QCD evolution of both the quark Sivers function and the fragmentation function 
as in Eqs.~\eqref{ff-form} and~\eqref{sivers-form} into Eq.~\eqref{fut}, we can eventually write 
$F_{UT}^{\sin\left(\phi_h -\phi_s\right)}$ as
\bea
F_{UT}^{\sin\left(\phi_h -\phi_s\right)} =& \frac{1}{4\pi z_h^2} \int_0^{\infty} 
db\, b^2 J_1(P_{h\perp}b/z_h) \sum_q e_q^2 \, T_{q, F}(x_B, x_B, c/b_*) D_{h/q}(z_h, c/b_*) 
\nnu
&\times
\exp\left\{-\int_{c^2/b_*^2}^{Q^2} \frac{d\mu^2}{\mu^2} \left(A\ln\frac{Q^2}{\mu^2}+B\right)\right\}
\exp\left\{-b^2\left(g_1^{\rm ff}+g_1^{\rm sivers}+ g_2\ln\frac{Q}{Q_0}\right)\right\},
\eea
with $J_1$ being the Bessel function of the first order. The Sivers asymmetry
$A_{UT}^{\sin(\phi_h-\phi_s)}$ is defined as
\bea
A_{UT}^{\sin(\phi_h-\phi_s)} = \frac{\sigma_0(x_B, y, Q^2)}{\sigma_0(x_B, y, Q^2)} 
\frac{F_{UT}^{\sin\left(\phi_h -\phi_s\right)}}{F_{UU}}.
\eea

If we want to use the above QCD formalism (with QCD evolution of TMDs included) 
to describe the Sivers asymmetries in SIDIS, we have to parametrize the usual Qiu-Sterman 
functions $T_{q,F}(x, x, \mu)$. For this part, following~\cite{Kouvaris:2006zy}, we assume they are 
proportional to the usual unpolarized collinear PDFs as
\bea
T_{q, F}(x, x, \mu) = N_q \frac{(\alpha_q+\beta_q)^{(\alpha_q+\beta_q)}}{\alpha_q^{\alpha_q} \beta_q^{\beta^q}} 
x^{\alpha_q} (1-x)^{\beta_q} f_{q/A}(x, \mu). 
\eea
We will have $\alpha_u, \alpha_d, N_u, N_d$ for $u$ and $d$ quarks, and $N_{\bar u}, 
N_{\bar d}, N_{s}, N_{\bar s}, \alpha_{\rm sea}$ for sea quarks. At the same time, we choose 
the same $\beta_q\equiv \beta$ for all quark flavors. Including $\langle k_{s\perp}^2\rangle 
= 4\,g_1^{\rm sivers}$ in the non-perturbative Sudakov factor Eq.~\eqref{sivers-form}, we have 
in total 11 fitting parameters.

\bef
\psfig{file=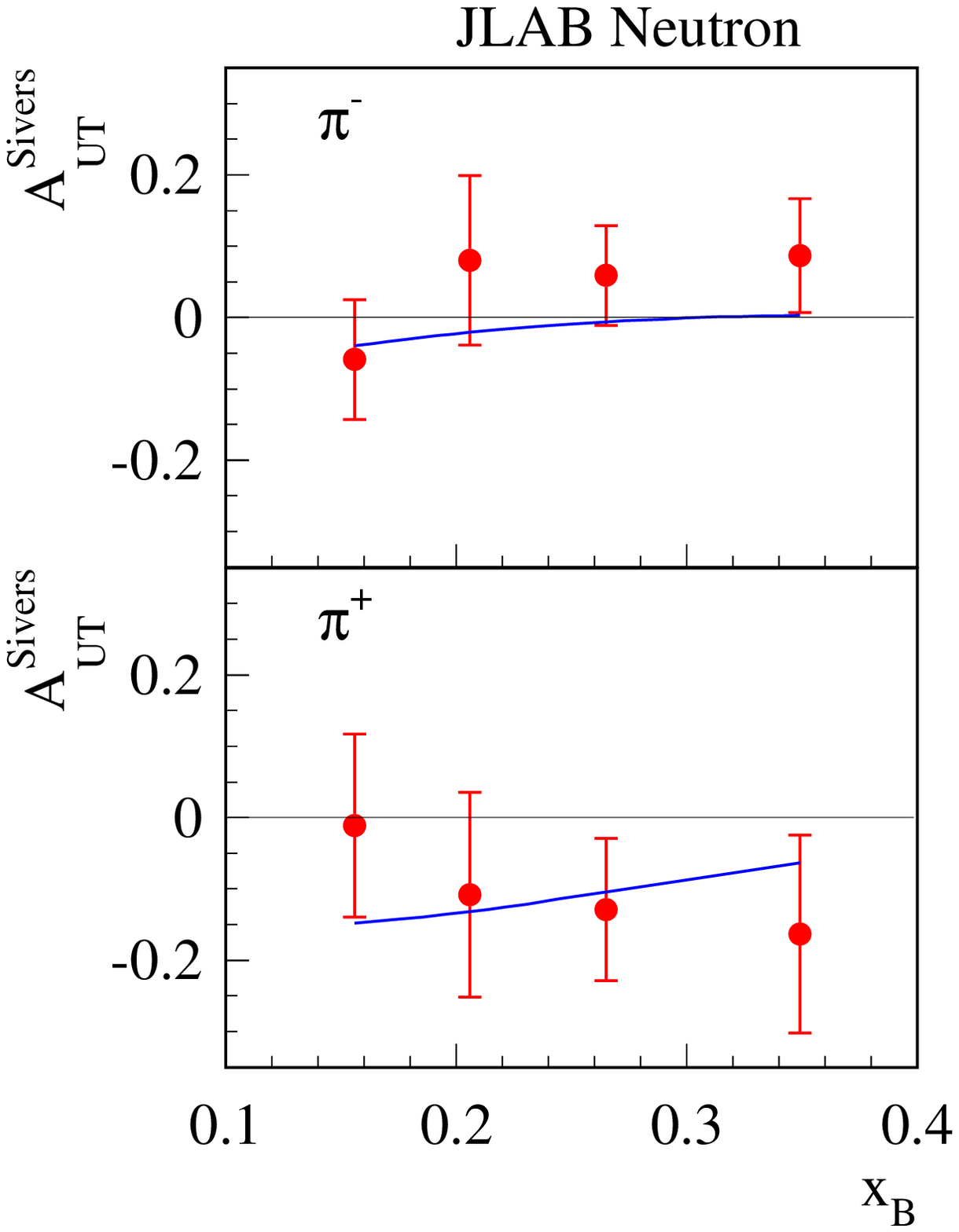, width=2in}
\caption{Results obtained from the TMD evolution fit of the SIDIS $A_{UT}^{\sin(\phi_h - \phi_s)}$ 
Sivers asymmetries are compared with the JLab experimental data~\cite{Qian:2011py} for charged 
pion production on a neutron target.}
\label{jlab}
\eef
\bef
\psfig{file=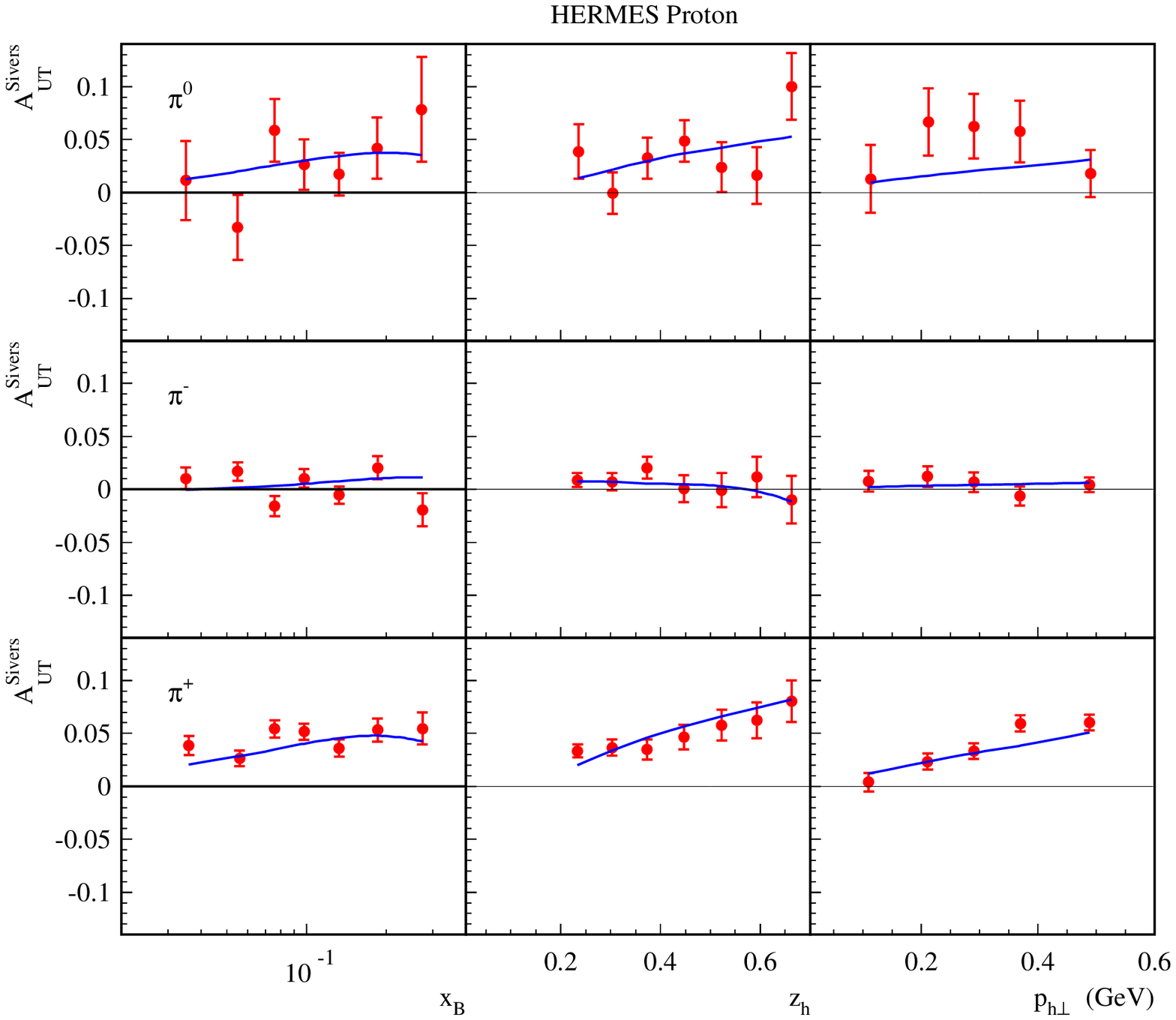, width=4.2in}
\caption{Results obtained from the TMD evolution fit of the SIDIS $A_{UT}^{\sin(\phi_h - \phi_s)}$ Sivers 
asymmetries are compared with the HERMES experimental data~\cite{Airapetian:2009ae} 
for neutral and charged pion production.}
\label{hermes-pion}
\eef
We use the MINUIT package to perform a global fit of the Sivers asymmetries data in SIDIS. 
To be consistent with the region of applicability of our QCD factorization formalism while still 
having enough experimental data in our analysis, we restrict our fit to the same transverse momentum 
region as specified in last section for the unpolarized differential cross section: 
for hadron production at JLab~\cite{Qian:2011py} with $\langle Q^2\rangle = 1.38 - 2.68$ GeV$^2$
we choose $P_{h\perp}\leq 0.5$ GeV; for hadron production at HERMES~\cite{Airapetian:2009ae} 
with $\langle Q^2\rangle \approx 2.45$ GeV$^2$, we choose $P_{h\perp}\leq 0.6$ GeV; and for 
the COMPASS experimental data~\cite{Alekseev:2008aa,Adolph:2012sp} with 
$\langle Q^2\rangle \approx 3 - 5$ GeV$^2$, we choose $P_{h\perp}\leq 0.7$ GeV. 
For the transversely polarized neutron and deuteron targets, we use  isospin symmetry to 
relate the quark Sivers functions to those in the proton target. By simultaneously fitting pion, 
kaon, and charged hadron experimental data~\cite{Airapetian:2009ae,Alekseev:2008aa,Adolph:2012sp,Qian:2011py} 
from JLab, HERMES, and COMPASS, we obtain an acceptable overall description of the experimental data 
with the total $\chi^2\approx 300$ for 241 data points, and thus $\chi^2/d.o.f. = 1.3$. 
The fitted parameters are given in the Table.~\ref{fitpar}. 
\begin{table}[htb]
\centering
\caption{Best values of the free parameters  for the Sivers function from our fit to 
SIDIS data~\cite{Airapetian:2009ae,Alekseev:2008aa,Adolph:2012sp,Qian:2011py} on $A_{UT}^{\sin(\phi_h-\phi_s)}$.}
\label{fitpar}
\begin{tabular}{l l l l l l}
\hline
\hline
&&$\chi^2/d.o.f. = 1.3$&&&\\
\hline
$\alpha_{u}$ &=& $1.051^{+0.192}_{-0.180}$  & $\alpha_{d}$ &=&  $1.552^{+0.303}_{-0.275}$ \\
$\alpha_{\rm sea}$ &=& $0.851^{+0.307}_{-0.305}$ & $\beta$ &=& $4.857^{+1.534}_{-1.395} $ \\
$N_{u}$ &=& $0.106^{+0.011}_{-0.009}$ & $N_{d}$ &=& $-0.163^{+0.039}_{-0.046} $ \\
$N_{\bar u}$ &=&  $-0.012^{+0.018}_{-0.020}$ & $N_{\bar d}$ &=&   $-0.105^{+0.043}_{-0.060}$\\
$N_{s}$ &=& $0.103^{+0.548}_{-0.604}$ & $N_{\bar s}$ &=& $-1.000{\pm 1.757}$ \\
$\langle k_{s\perp}^2\rangle$ &=& $0.282^{+0.073}_{-0.066}$ GeV$^2$ & & & \\
\hline
\hline
\end{tabular}
\end{table}

Comparison of the fits to the experimental data are presented in Figs.~\ref{jlab} - \ref{compass-hadron}, 
with the solid curves representing our fitted theoretical results.  In Fig.~\ref{jlab} we show the 
comparison with the JLab experimental data~\cite{Qian:2011py} for charged pion production on a 
neutron target. JLab experimental data have a relatively large error bar for the asymmetries and have 
only the $x_B$-dependence of the Sivers asymmetries. On the other hand, both HERMES and COMPASS 
experimental data have the Sivers asymmetries as functions of $x_B$, $z_h$, and $P_{h\perp}$, respectively. 
In Figs.~\ref{hermes-pion} and \ref{hermes-kaon} we show the results obtained from our fit 
compared with the HERMES experimental data~\cite{Airapetian:2009ae}  for pion and kaon production 
on a proton target, respectively. In Figs.~\ref{compass-pion} and~\ref{compass-kaon} we present 
the comparison with the COMPASS experimental data for charged pion and kaon production on a deuteron 
target~\cite{Alekseev:2008aa}. Finally, in Fig.~\ref{compass-hadron} we show the comparison with 
the COMPASS experimental data for charged hadron production on a proton target~\cite{Adolph:2012sp}. 
One  sees that the fit is of  rather good quality. Even thought the $\chi^2/d.o.f.$ is slightly larger 
than earlier Gaussian-form  fits for the TMDs \cite{Anselmino:2008sga}, we feel more confident about 
our results as they are  based on a QCD formalism which can give a rather good description for 
all the corresponding unpolarized differential cross sections. 
\bef
\psfig{file=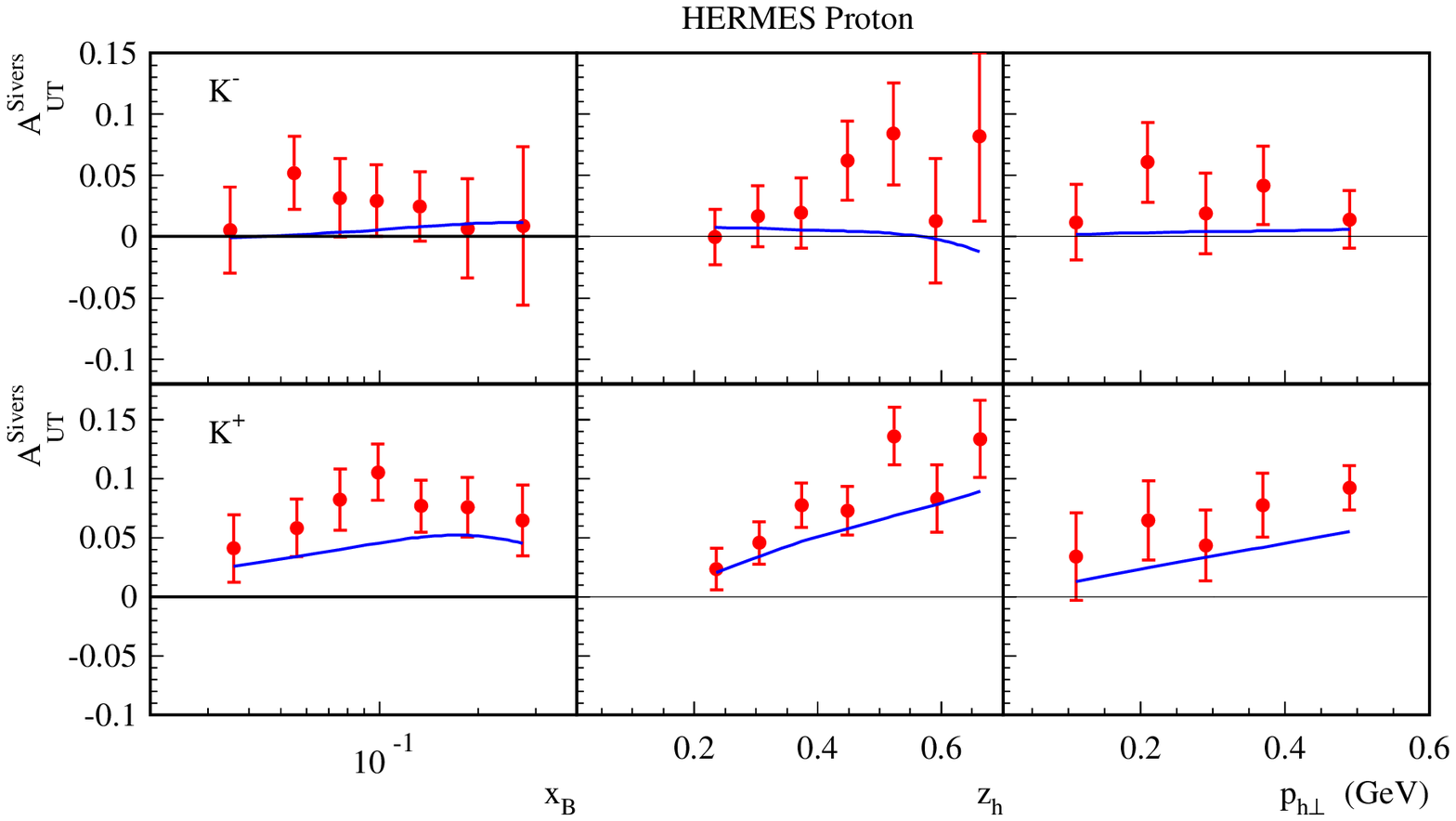, width=4.2in}
\caption{Results obtained from the TMD evolution fit of the SIDIS $A_{UT}^{\sin(\phi_h - \phi_s)}$ 
Sivers asymmetries are compared with the HERMES experimental data \cite{Airapetian:2009ae} 
for kaon production.}
\label{hermes-kaon}
\eef
\bef
\psfig{file=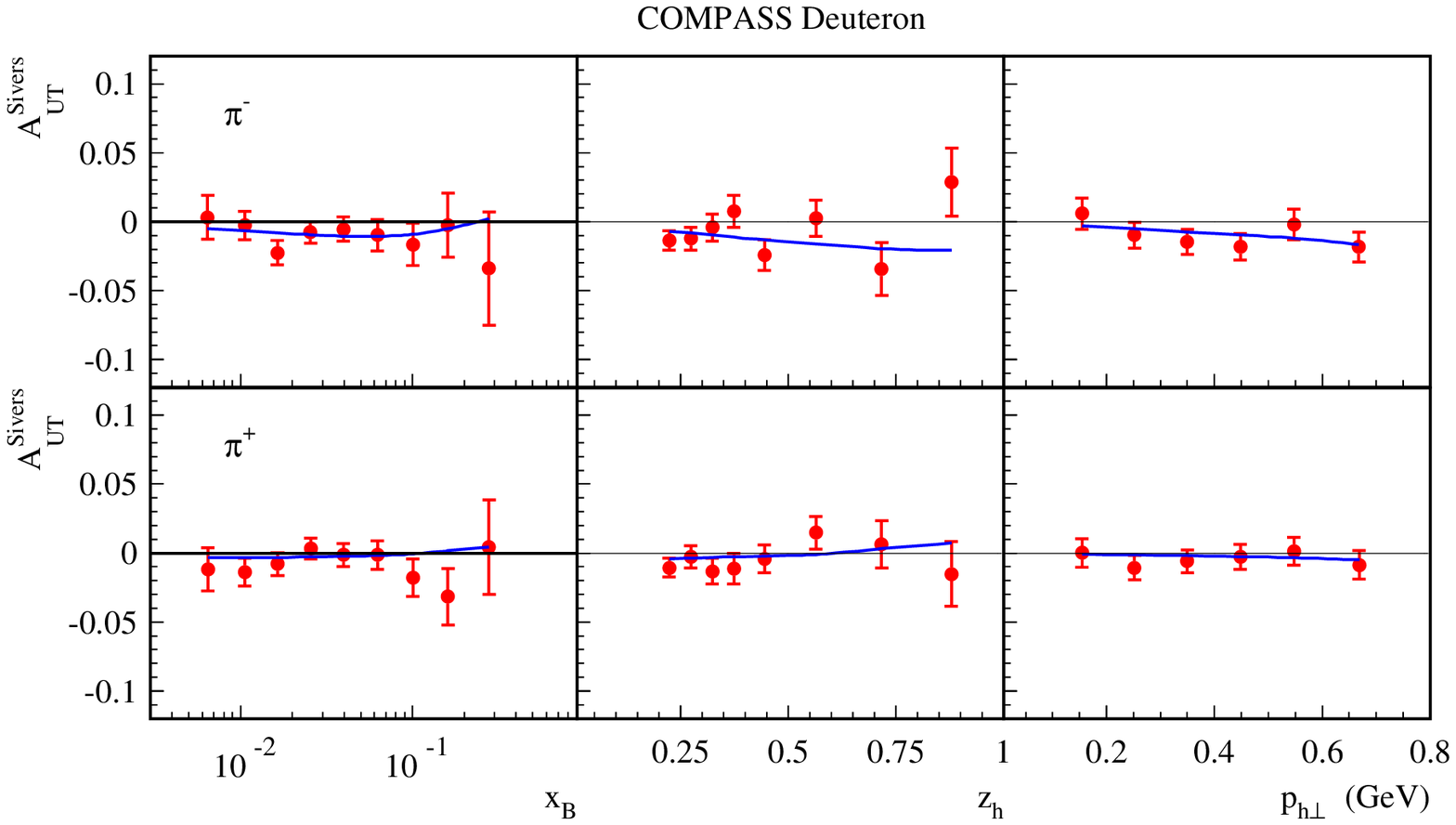, width=4.2in}
\caption{Results obtained from the TMD evolution fit of the SIDIS $A_{UT}^{\sin(\phi_h - \phi_s)}$ 
Sivers asymmetries are compared with the COMPASS experimental data \cite{Alekseev:2008aa} for 
charged pion production on a Deuteron target.}
\label{compass-pion}
\eef
\bef
\psfig{file=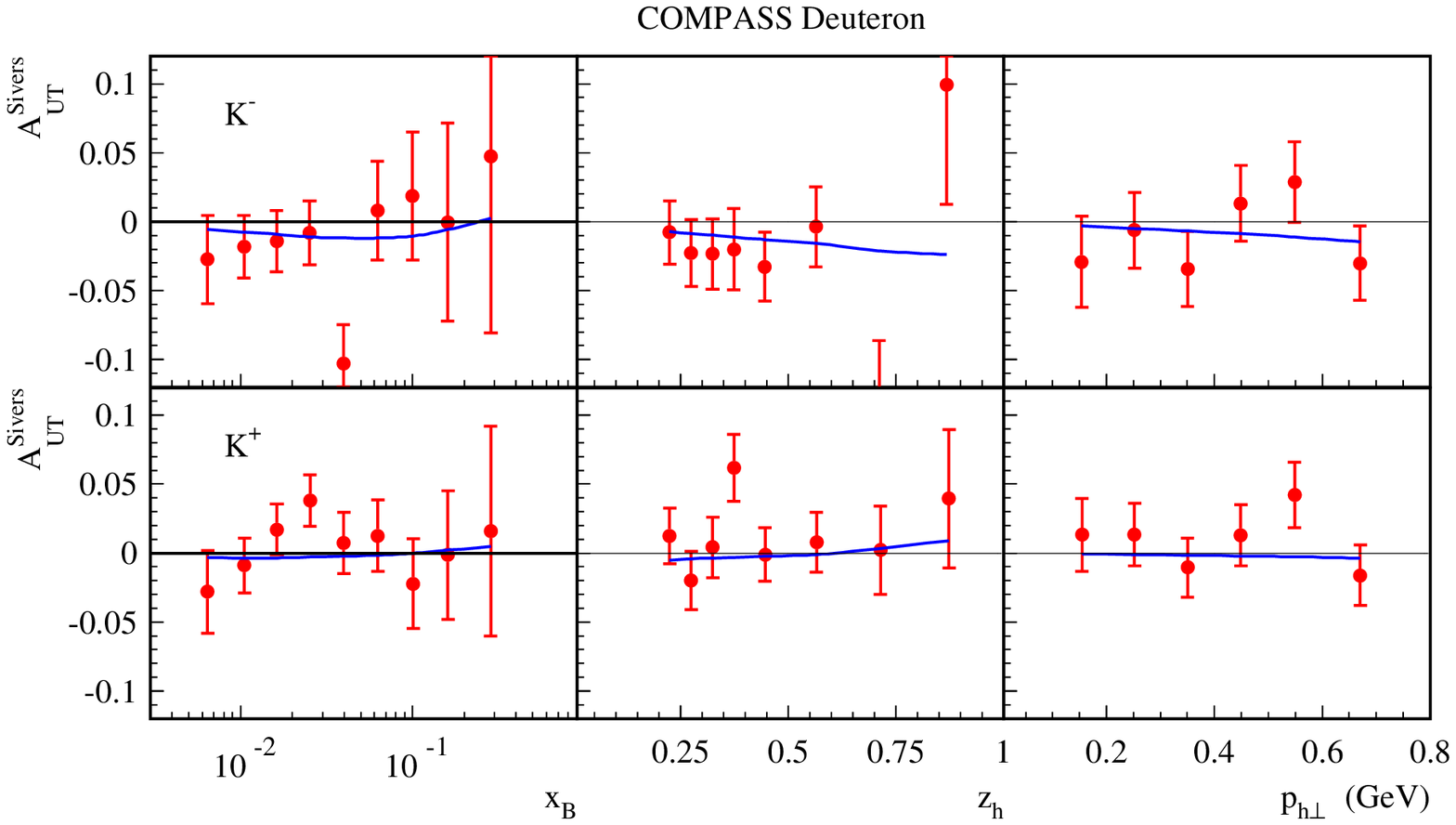, width=4.2in}
\caption{Results obtained from the TMD evolution fit of the SIDIS $A_{UT}^{\sin(\phi_h - \phi_s)}$ 
Sivers asymmetries are compared with the COMPASS experimental data \cite{Alekseev:2008aa} 
for kaon production on a Deuteron target.}
\label{compass-kaon}
\eef
\bef
\psfig{file=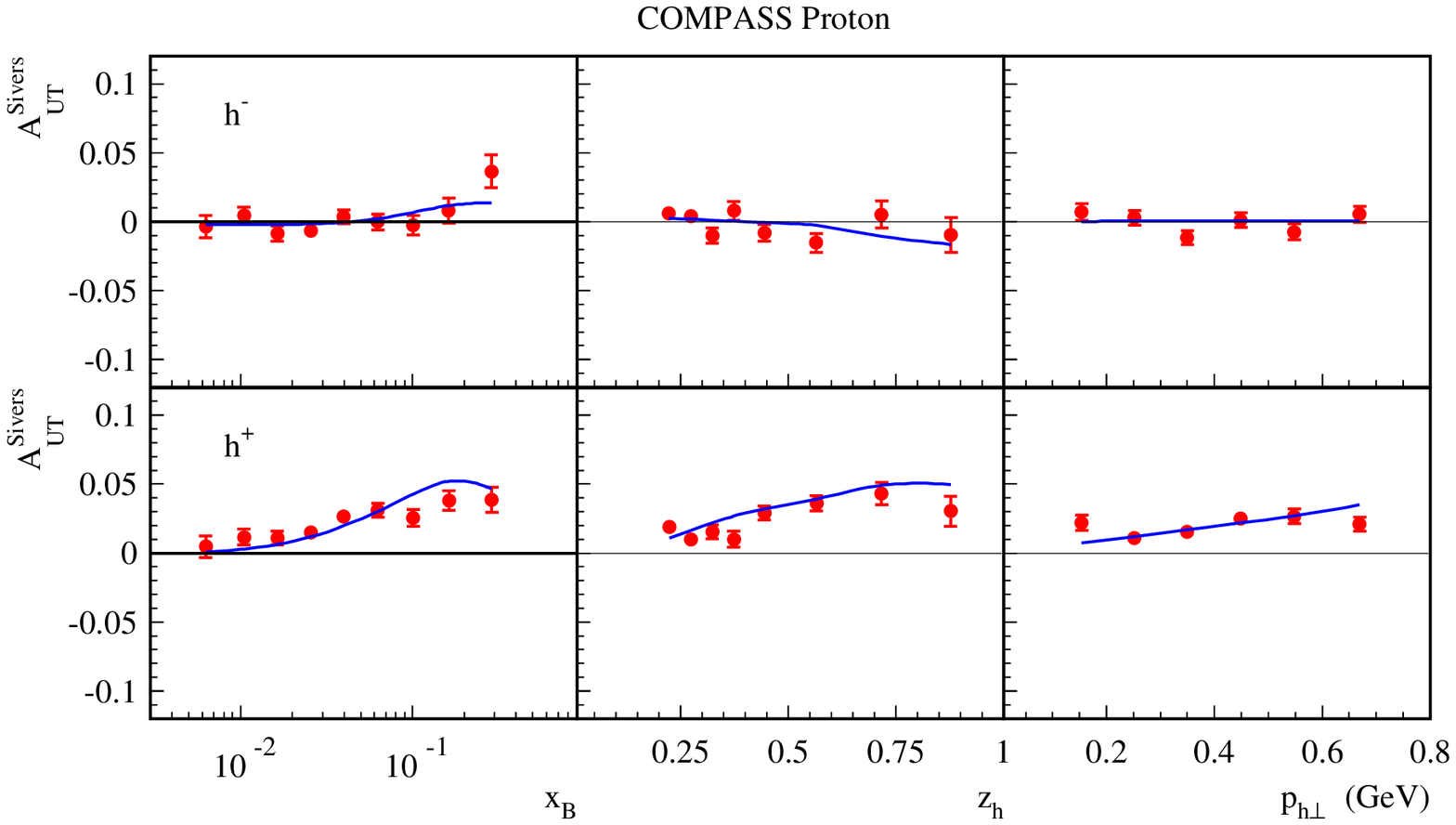, width=4.2in}
\caption{Results obtained from the TMD evolution fit of the SIDIS $A_{UT}^{\sin(\phi_h - \phi_s)}$ 
Sivers asymmetries are compared with the COMPASS experimental data \cite{Adolph:2013stb} 
for charged hadron production on a proton target.}
\label{compass-hadron}
\eef

In Fig.~\ref{qiu-sterman-fit},  we show the Qiu-Sterman function $T_{q,F}(x, x, Q)$ extracted
from our fits for $u$, $d$, and $s$ quark flavors as a function of parton momentum fraction $x$ 
at a scale $Q^2=2.4$ GeV$^2$. We find $T_{u,F}(x, x, Q)$ and $T_{d,F}(x, x, Q)$ have a similar 
size but opposite sign, which is consistent with previous extractions from the SIDIS 
process~\cite{Anselmino:2008sga, Bacchetta:2011gx, Sun:2013hua}. The current extraction with the limited kinematic 
coverage from the experimental data can only constrain reasonably well the $u$ and $d$ quark Sivers 
functions. All of the sea quark Sivers functions are not constrained well. For example, even 
if we neglect all the sea quark Sivers functions in our formalism, we obtain a similar 
$\chi^2/d.o.f.$. In this respect, the future planned electron-ion collider experiments and 
the DY and $W$ boson production \cite{Aschenauer:2013woa, DY-compass-exp, DY-fermi-beam, DY-fermi-target, DY-rhic-exp} 
should provide us with 
better constraints on the sea quark Sivers functions.
\bef
\psfig{file=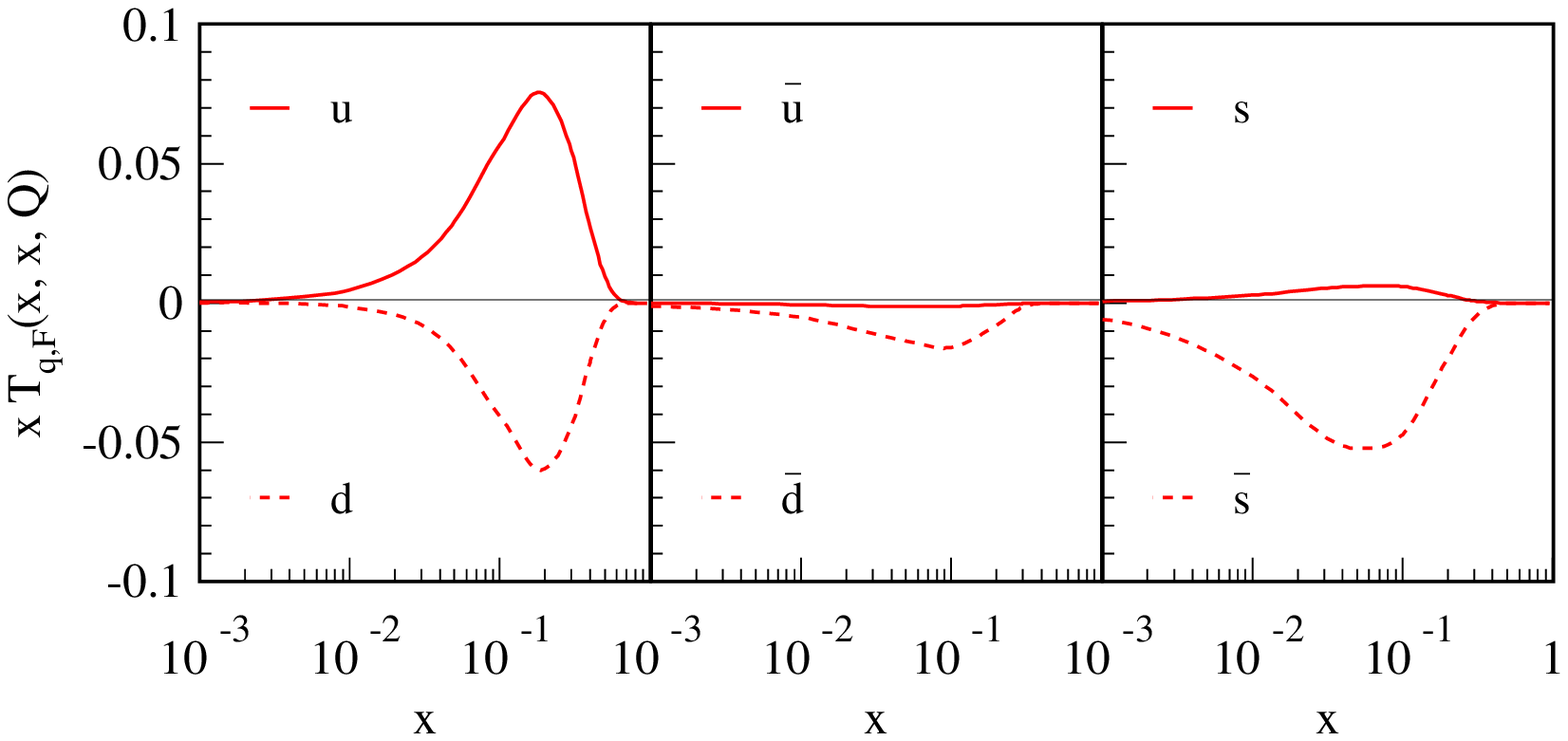, width=4.2in}
\caption{Qiu-Sterman function $T_{q,F}(x, x, Q)$ for $u$, $d$, and $s$ flavors at a 
scale $Q^2=2.4$ GeV$^2$, as extracted by our simultaneous fit of JLab, HERMES, and COMPASS data.}
\label{qiu-sterman-fit}
\eef

\subsection{Predictions for the Sivers effect in DY production}
One of the most important properties of the Sivers function is its time-reversal 
modified universality, which has been extensively studied in recent years. In particular, 
the Sivers function changes sign while keeping its magnitude  when going from the SIDIS process 
to the DY processes. Testing this sign change has become one of the hot topics in hadron physics 
in recent years. There have been calculations for the Sivers asymmetries in DY production 
based on the naive parton model without QCD evolution of the TMDs, 
see~\cite{Kang:2009sm,Collins:2005rq,Vogelsang:2005cs,Anselmino:2009st}. One of the most 
recent papers~\cite{Sun:2013hua} has taken into account the QCD evolution of TMDs, in which 
the authors use a different Sudakov factor for the evolution in the region from the low $Q$ 
to the intermediate $Q\sim 10$ GeV, and in the region from the intermediate $Q\sim 10$ GeV 
to high $Q\sim M_{W/Z}$.  In this section, we will use the information on the Sivers 
functions obtained through our fit to make predictions for the Sivers asymmetries for 
both DY lepton pair and $W$ boson production in $pp$ collisions. Importantly, we are 
able to use the same universal Sudakov factor in the QCD evolution for the whole $Q$ region: 
from low $Q$ up to high $Q\sim M_{W/Z}$.

For Drell-Yan production in single transversely polarized $p^\uparrow p$ collisions, 
$A^\uparrow(P_A, s_\perp) + B(P_B) \to [\gamma^*\to] \ell^+\ell^-(y, Q, q_\perp)+X$, the unpolarized 
differential cross section at small $p_\perp\ll Q$ is given by Eq.~\eqref{DY}, while the spin-dependent 
cross section $\Delta \sigma \equiv \left[\sigma(s_\perp)-\sigma(-s_\perp)\right]/2$ can be written 
as~\cite{Kang:2011mr,Kang:2012vm}
\bea
\frac{d\Delta\sigma}{dQ^2 dy d^2p_\perp} &= \epsilon^{\alpha\beta} s_\perp^\alpha 
\sigma_0^{\rm DY} \int \frac{d^2b}{(2\pi)^2} e^{ip_\perp\cdot b} \sum_q e_q^2\, 
f_{1T, \rm DY}^{\perp,q(\beta)}(x_a, b; Q)  f_{\bar q/B}(x_b, b; Q),
\nnu
&= - \frac{\sigma_0^{\rm DY}}{4\pi} \int_0^{\infty} db\, b^2J_1(p_\perp b) 
\sum_q e_q^2\, T_{q, F}(x_a, x_a, c/b^*) 
f_{\bar q/B}(x_b, c/b^*)  
\nnu
&\times
\exp\left\{-\int_{c/b^*}^Q \frac{d\mu^2}{\mu^2} \left(A\ln\frac{Q^2}{\mu^2}+B\right)\right\} 
\exp\left\{-b^2\left(g_1^{\rm pdf}+g_1^{\rm sivers}+g_2\ln\frac{Q}{Q_0}\right)\right\}.
\label{spin-dy}
\eea
To arrive at the second expression in Eq.~\eqref{spin-dy}, we first apply the sign change 
for the Sivers functions between the SIDIS and the DY processes
\bea
f_{1T, \rm DY}^{\perp,q(\beta)}(x_a, b; Q)  = - f_{1T, \rm SIDIS}^{\perp,q(\beta)}(x_a, b; Q).
\eea
We then use Eq.~\eqref{sivers-form} and Eq.~\eqref{spin-dy} and  follow the experimental 
convention to choose the pair's transverse momentum $p_\perp$ along the  $x$-direction, while 
the spin vector $s_\perp$ is along $y$-direction~\cite{Kang:2011hk,Anselmino:2009pn} and the transversely 
polarized proton is moving in the $+z$-direction. The single transverse spin asymmetry for DY production is given by
\bea
A_N = \left.\frac{d\Delta\sigma}{dQ^2 dy d^2p_\perp}\right/ \frac{d\sigma}{dQ^2 dy d^2p_\perp}.
\eea
It is important to realize that the $A_N$ defined above is opposite to the so-called weighted asymmetry 
$A_N^{\sin(\phi_{\gamma}- \phi_s)}$ defined in the literature, see, e.g., Refs.~\cite{Anselmino:2009st,Kang:2009sm}. 

\bef
\psfig{file=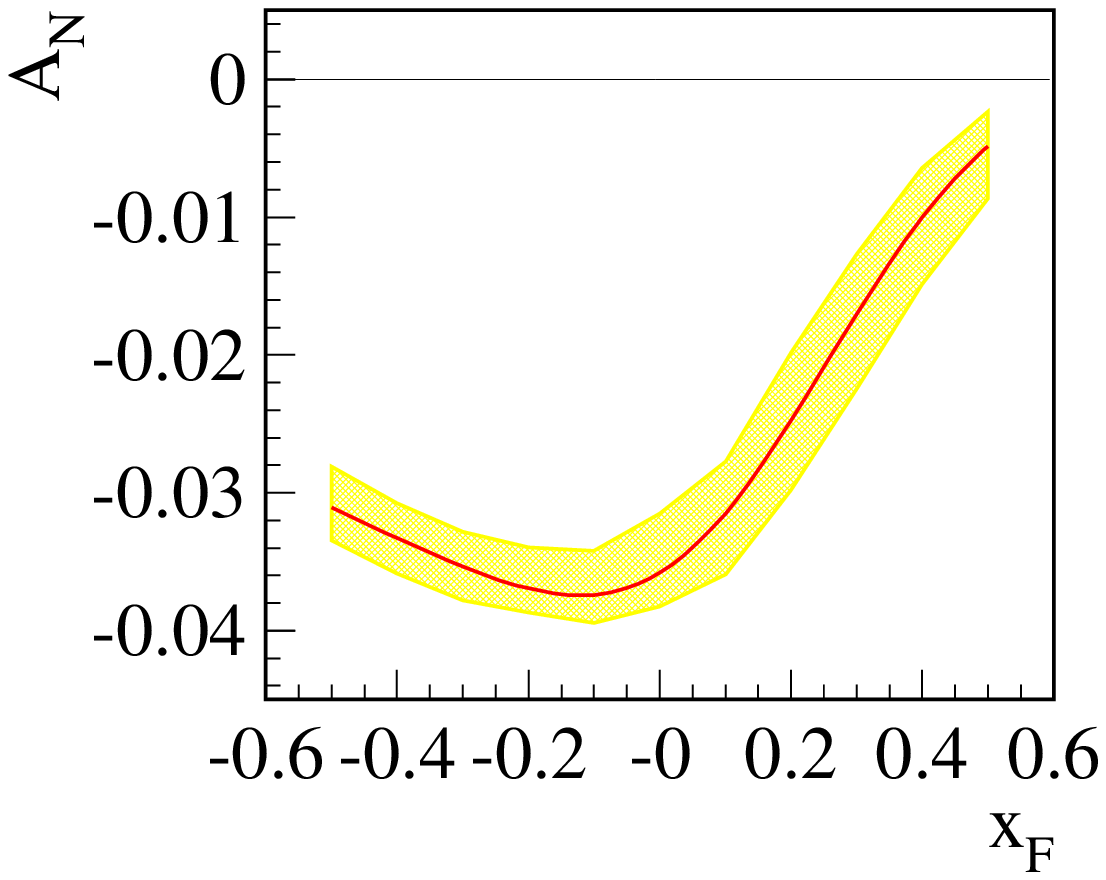, width=2.2in}
\hskip 0.1in
\psfig{file=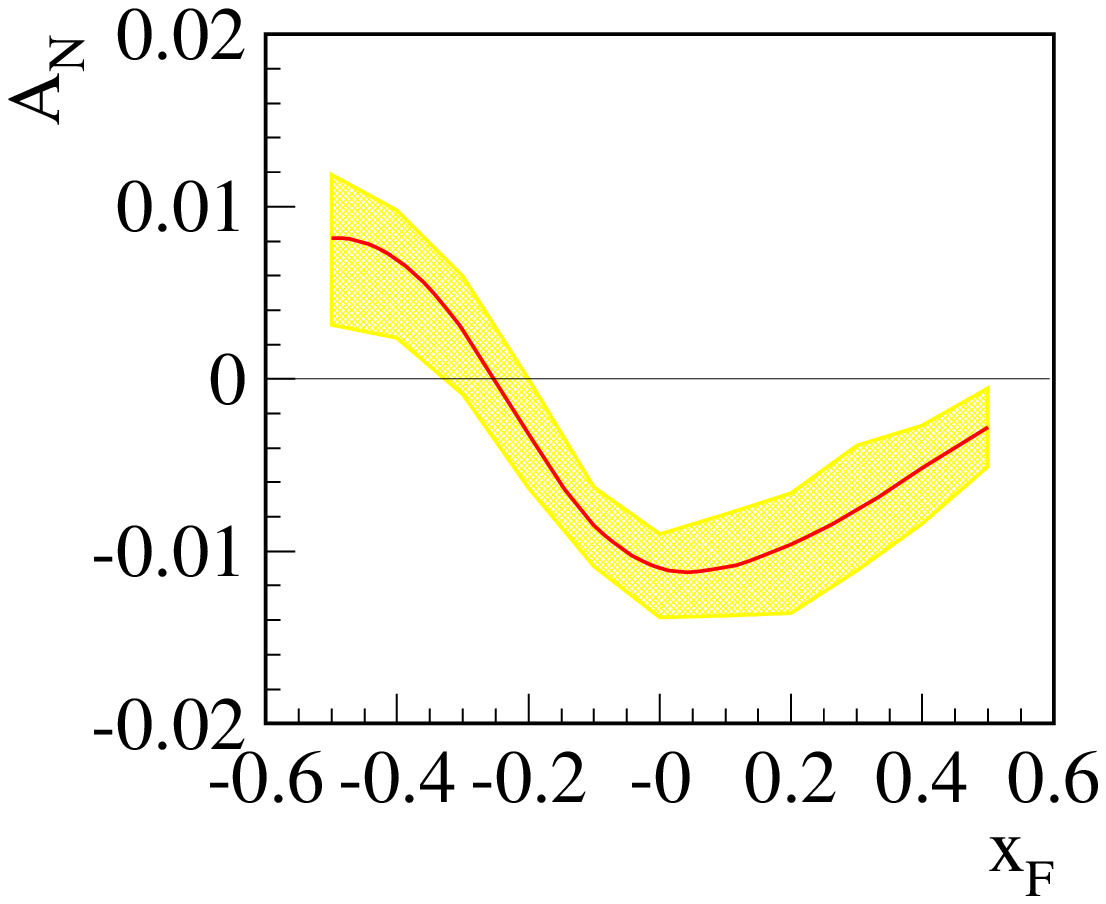, width=2.2in}
\hskip 0.1in
\psfig{file=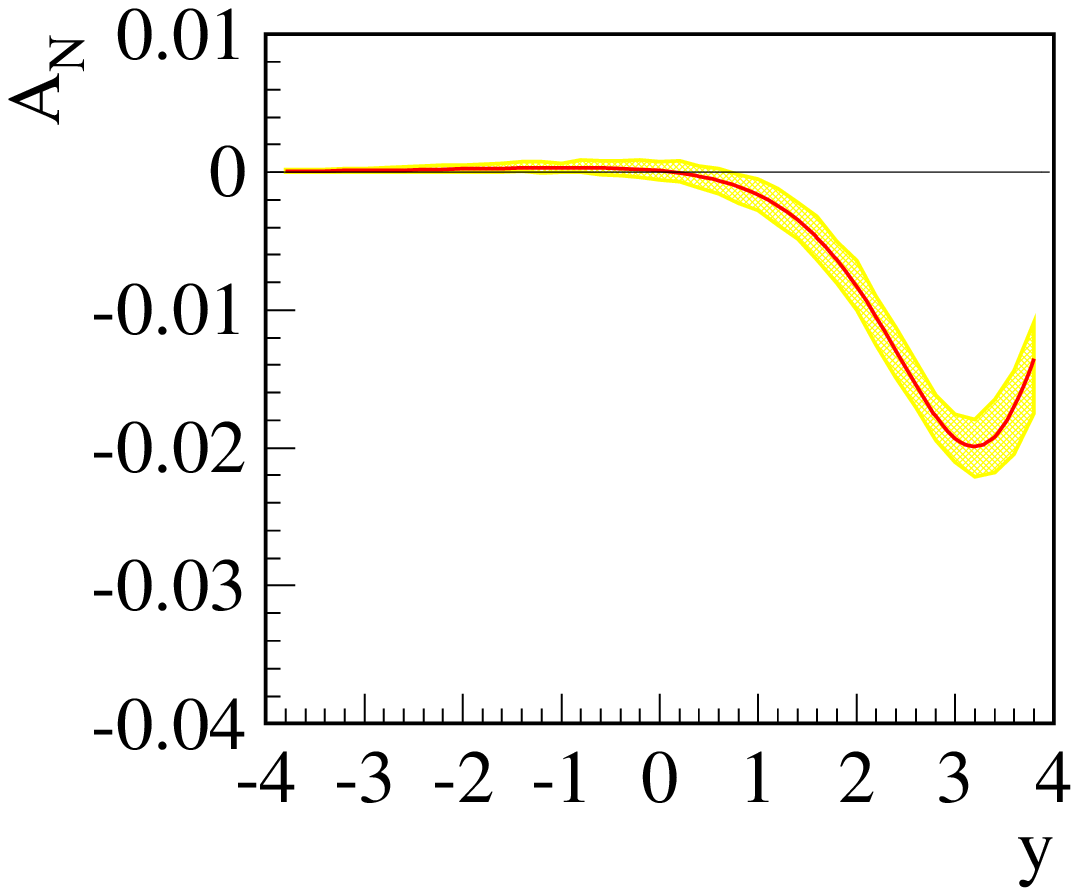, width=2.2in}
\caption{Estimated Sivers asymmetries for DY lepton pair production. Left plot: $A_N$ in $p^\uparrow \pi^-$ collisions 
as a function of $x_F$ at COMPASS energy $\sqrt{s}=18.9$ GeV. Middle plot: $A_N$ in $p^\uparrow p$ collisions is plotted as a 
function of $x_F$ at Fermilab energy $\sqrt{s}=15.1$ GeV. Right plot: $A_N$ in $p^\uparrow p$ collisions is plotted 
as a function of the pair's rapidity $y$ at RHIC energy $\sqrt{s}=510$ GeV. We have integrated over
the pair's transverse momentum $0<p_\perp<1$ GeV in the invariant mass range $4<Q<9$ GeV.}
\label{DY-sivers}
\eef
There are several planned experiments to measure the $A_N$ for DY lepton pair production. The COMPASS collaboration at 
CERN will use a 190~GeV $\pi^-$ beam to scatter on the polarized proton target~\cite{DY-compass-exp}, 
which corresponds to a CM energy $\sqrt{s}=18.9$ GeV. At Fermilab, one can use the 120 GeV proton beam in the main injector. 
There are two proposals corresponding to either a polarized proton beam~\cite{DY-fermi-beam} or a polarized 
proton target~\cite{DY-fermi-target}. In both cases, the CM energy is $\sqrt{s}=15.1$ GeV. Finally, a DY measurement is also 
planned at RHIC~\cite{Aschenauer:2013woa, DY-rhic-exp}. In the following, we will present an estimate of the  Sivers asymmetry 
based on our evolution approach. For better comparison, we will always present the asymmetry in the center-of-mass frame 
of the colliding particles. We further choose the transversely polarized proton to move in the $+z$ direction, while 
the other unpolarized particle ($\pi^-$ for COMPASS and the unpolarized proton for Fermilab and RHIC) moves 
in the $-z$ direction. We define 
\bea
x_F = x_a - x_b,
\label{xf}
\eea
which is the Feynman-$x$ at tree level with $x_{a,b}$ given by Eq.~\eqref{xab}. Here, $x_a$ is always the parton momentum 
fraction in the transversely polarized proton, while $x_b$ is the parton momentum fraction in the other unpolarized particle. It 
is important to mention that these conventions could differ from those used in some experiments~\cite{DY-compass-exp, DY-fermi-target}.

\bef
\psfig{file=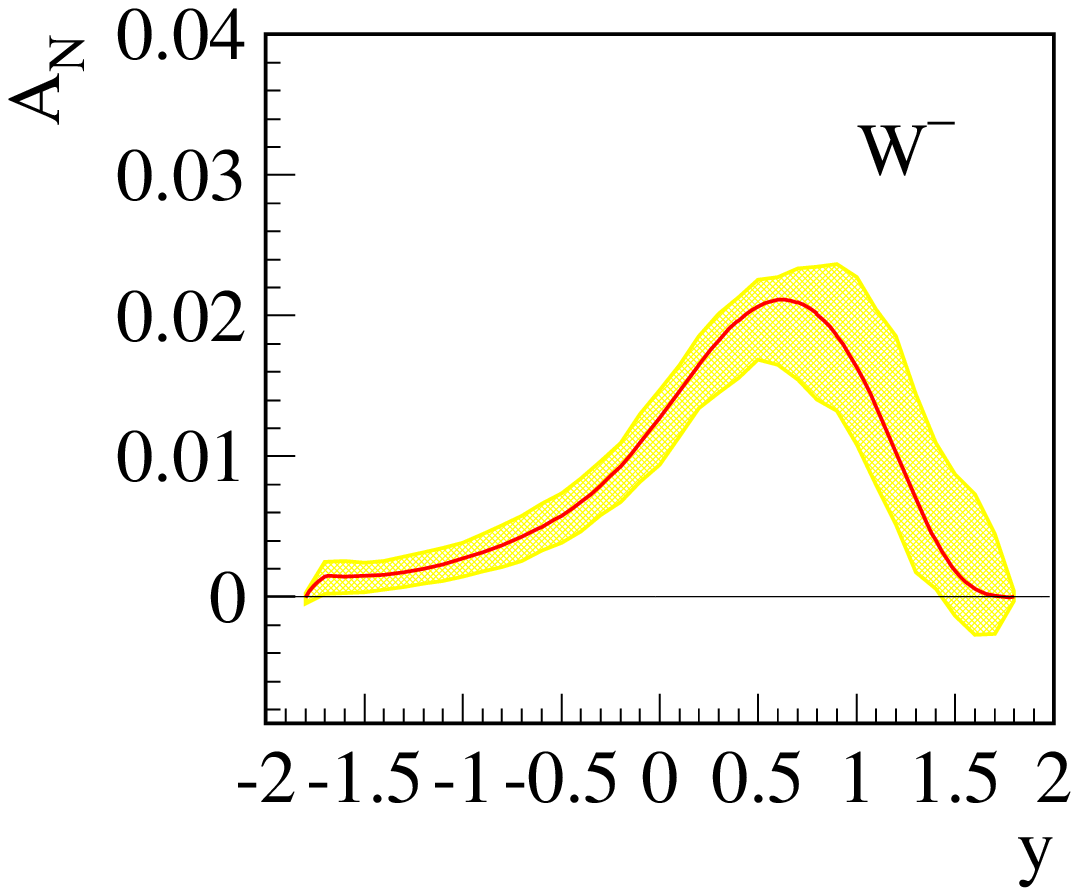, width=2.2in}
\hskip 0.1in
\psfig{file=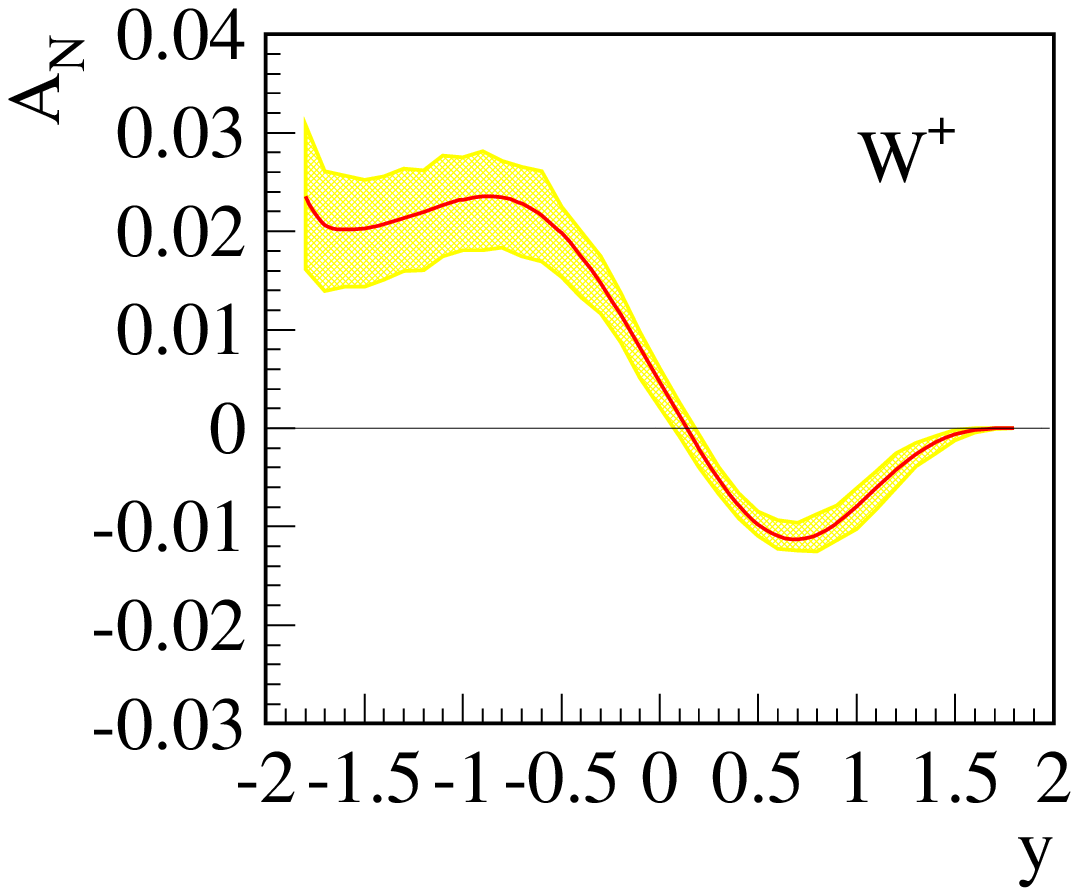, width=2.2in}
\caption{Estimated Sivers asymmetries as a function of rapidity $y$ for $W^-$  and $W^+$ production at the  
RHIC energy $\sqrt{s}=510$ GeV. We have integrated over the 
transverse momentum for $W$ boson in $0<p_\perp <3$ GeV.}
\label{Wboson}
\eef
In Fig.~\ref{DY-sivers} (left) we plot our predicted Sivers asymmetry $A_N$ for DY lepton pair 
production as a function of $x_F$ for COMPASS kinematics $\sqrt{s}=18.9$ GeV. For the pion beam, 
we use the PDFs in the pion extracted in~\cite{Sutton:1991ay}. We have 
integrated over the transverse momentum $0<p_\perp<1$~GeV and invariant mass of the pair $4<Q<9$~GeV. 
The solid curve corresponds to the calculation based on the best fit for the parameters in Table.~\ref{fitpar}, 
while the shaded area in the figure corresponds to the 1$\sigma$ error in the fitted parameters. 
COMPASS projected their measurement around $x_{\pi} - x_{p^\uparrow}\approx 0.2$ (corresponding to our $x_F=-0.2$ in 
Eq.~\eqref{xf})~\cite{DY-compass-exp}. The estimated asymmetry is around $3-4\%$ and should be measurable. 

In Fig.~\ref{DY-sivers} (middle) we plot the estimated Sivers asymmetry for the Fermilab energy 
$\sqrt{s}=15.1$~GeV. The proposed ``polarized beam'' experiment~\cite{DY-fermi-beam} will correspond 
to the region $0<x_F<0.6$, while the proposed ``polarized target'' experiment~\cite{DY-fermi-target} 
will roughly correspond to the region $-0.6<x_F<0.1$ in our notation Eq.~\eqref{xf}. The asymmetry 
is around $1-2\%$, which we hope it could be measured in the future. Finally, in Fig.~\ref{DY-sivers} (right) 
we plot $A_N$ as a function of the pair's rapidity $y$ at RHIC energy $\sqrt{s}=510$ GeV. We 
find that the asymmetry is around $2-3\%$ 
in the forward rapidity, which should be measurable at RHIC. 

Both the $u$ and $d$ quark Sivers functions contribute to the Sivers asymmetries in DY lepton 
pair production in $pp$ collisions. Since $u$ and $d$ quark Sivers functions have opposite sign, 
as shown in last subsection, 
they partially cancel each other in their contribution to the DY asymmetry. In order 
to be able to test the sign change of the quark Sivers function separately, $W$ boson asymmetries 
have been proposed~\cite{Kang:2009bp} and have been planned at RHIC experiment~\cite{Aschenauer:2013woa}. 
In Fig.~\ref{Wboson}, we plot our predicted Sivers asymmetries $A_N$ as a 
function of rapidity $y$ for $W^-$ and $W^+$ boson production, respectively. The transverse momentum 
is integrated  over $0<p_\perp < 3$ GeV and $\sqrt{s}=510$ GeV. The $W^-$ asymmetry at forward rapidity 
is sensitive to the $d$-quark Sivers function. On the other hand, the $W^+$ asymmetry is sensitive to 
$u$ quark Sivers function at forward rapidity, while it receives contributions from both 
the $\bar{d}$ and $\bar{s}$ quark  Sivers functions in the backward rapidity region. 
As we emphasized in the last subsection, sea quark Sivers functions are not constrained well by the 
current SIDIS data. Thus the future DY and $W$ boson asymmetry measurements should provide valuable 
information on the sea quark Sivers functions. The $W$ boson asymmetry can be quite large if calculated 
in a naive parton model without QCD evolution~\cite{Kang:2009bp}. Once the QCD evolution is taken 
into account, the asymmetry is only about $2 - 3\%$. We hope it can still be measured by the 
RHIC experiments. 

\section{Summary}
In this paper we studied the QCD evolution of Sivers asymmetries in both semi-inclusive deep 
inelastic scattering (SIDIS) and Drell-Yan production (DY). Since QCD evolution of TMDs involves 
both perturbative and non-perturbative parts, we verified that the non-perturbative part of the 
evolution kernel plays a very important role for phenomenological studies. Consequently, we 
placed special emphasis on the non-perturbative Sudakov factor in the evolution formalism. 
Since one essential part of this Sudakov factor is spin-independent, we first found a form which 
can describe reasonably well the experimental data for the transverse momentum distribution in SIDIS at relatively 
low momentum scale $Q$, DY lepton pair production at intermediate $Q$, and $W/Z$ production 
at high $Q$. Once this part of the QCD evolution was fixed, we then used the same 
Sudakov factor to perform a global analysis  of all the experimental data on the Sivers asymmetry 
from HERMES, COMPASS, and Jefferson Lab. We extracted the quark Sivers functions in SIDIS 
from such a global fitting procedure and used them with a reversed  sign to make predictions 
for the Sivers asymmetries for DY lepton pair and $W$ production. We found that the valence quark region 
is well-constrained by existing measurements but the sea quark asymmetry cannot be reliably determined.  
Hence, it is important that these predictions  be compared with the experimental measurements in 
the near future to not only test the sign change of the Sivers effect but also to determine 
much more accurately the Sivers functions for sea quarks.

\section*{Acknowledgments}
M.G.E. and A.I. thank I. Scimemi for exchanging some thoughts regarding the content of this work compared with that of Ref. \cite{madridtorino}. 
Z.K. thanks E.~C.~Aschenauer, A.~Bacchetta, D.~Boer, M.~Boglione, L.~Gamberg, A.~Metz, A.~Prokudin, J.~ Qiu, A.~Signori, and W.~Vogelsang for helpful discussions and useful comments. 
This research is supported by the 
US Department of Energy, Office of Science and in part by the LDRD program at LANL.
It is also part of the research program of the ``Stichting voor Fundamenteel Onderzoek der Materie (FOM)'', which is financially supported by the ``Nederlandse Organisatie voor Wetenschappelijk Onderzoek (NWO)'', the US Department of Energy under Grant No.~DE-SC0008745 and the Spanish MECD, Grant No.~FPA2011-27853-CO2-02.


\end{document}